\documentclass[twocolumn,numberedappendix,appendixfloats,twocolappendix]{openjournal}

\usepackage{lipsum}

\usepackage[usenames,dvipsnames]{xcolor}
\usepackage{textgreek}
\usepackage[utf8]{inputenc}
\usepackage[english]{babel}

\usepackage{hyperref}

\hypersetup{
    unicode, 
    colorlinks=true,
    linkcolor=linkcolor,
    citecolor=linkcolor,
    filecolor=linkcolor,
    urlcolor=linkcolor,
}
\usepackage{color,colortbl}
\definecolor{linkcolor}{RGB}{28, 96, 214}
\usepackage{tensind}
\tensordelimiter{?}
\DeclareGraphicsExtensions{.bmp,.png,.jpg,.pdf}
\usepackage{verbatim}
\usepackage[normalem]{ulem}
\usepackage{orcidlink}
\usepackage{soul}

\usepackage{newtxtext,newtxmath}
\usepackage{amsmath,amsfonts,amssymb,slashed,float,bm,algorithmic,bbold,wasysym}
\usepackage{array,multirow,multirow,dcolumn,booktabs}
\usepackage{physics}
\usepackage{xspace}
\usepackage{tikz} 

\usepackage{lineno}

\begingroup
\lccode`\~=`\|
\lowercase{\endgroup
  \def~}{\,\middle|\,}
\mathcode`|="8000 

\NewDocumentCommand{\Prob}{m}
{
  \ensuremath{\operatorname{P}\left( #1 \right)}
}

\usepackage[capitalize]{cleveref}
\crefformat{footnote}{footnote~#2#1#3}
\usepackage{aas_macros}

\makeatletter
\usepackage{etoolbox}
\patchcmd\H@refstepcounter{\protected@edef}{\protected@xdef}{}{}
\makeatother

\DeclareUnicodeCharacter{2212}{-}

\newcommand{\bth}{\bm{\theta}}
\newcommand{\vd}{{\vb{x}}}
\newcommand{\vt}{{\vb{t}}}
\newcommand{\vdp}{{\vb{y}}}
\newcommand{\vdobs}{{\vb{x}_{\rm obs}}}
\newcommand{\vdpobs}{{\vb{y}_{\rm obs}}}
\newcommand{\ie}{{i.e.}\xspace}
\newcommand{\vs}{{\it vs}\xspace}

\graphicspath{ {./figs/} }

\begin{document}

\begin{nolinenumbers}
\vspace*{-\headsep}\vspace*{\headheight}
\footnotesize \hfill FERMILAB-PUB-26-0659-PPD\\
\vspace*{-\headsep}\vspace*{\headheight}
\footnotesize \hfill DES-2025-0940
\end{nolinenumbers}

\title{Dark Energy Survey Year 6 Results: fast and interpretable posterior predictive checks for correlated cosmic probes}


\author{C.~Doux~\orcidlink{0000-0003-4480-0096}}\affiliation{Department of Physics and Astronomy, University of Pennsylvania, Philadelphia, PA 19104, USA}
\affiliation{Universit\'e Grenoble Alpes, CNRS, LPSC-IN2P3, 38000 Grenoble, France}

\author{J.~Muir~\orcidlink{0000-0002-7579-770X}}\affiliation{Department of Physics, University of Cincinnati, Cincinnati, Ohio 45221, USA}
\affiliation{Perimeter Institute for Theoretical Physics, 31 Caroline St. North, Waterloo, ON N2L 2Y5, Canada}

\author{A.~Fert\'e~\orcidlink{0000-0003-3065-9941}}\affiliation{SLAC National Accelerator Laboratory, Menlo Park, CA 94025, USA}

\author{M.~Raveri~\orcidlink{0000-0002-7354-3802}}\affiliation{Department of Physics, University of Genova and INFN, Via Dodecaneso 33, 16146, Genova, Italy}

\author{D.~Sanchez Cid~\orcidlink{0000-0003-3054-7907}}\affiliation{Centro de Investigaciones Energ\'eticas, Medioambientales y Tecnol\'ogicas (CIEMAT), Madrid, Spain}
\affiliation{Physik-Institut, University of Zürich, Winterthurerstrasse 190, CH-8057 Zürich, Switzerland}

\author{N.~Weaverdyck~\orcidlink{0000-0001-9382-5199}}\affiliation{Berkeley Center for Cosmological Physics, Department of Physics, University of California, Berkeley, CA 94720, US}
\affiliation{Lawrence Berkeley National Laboratory, 1 Cyclotron Road, Berkeley, CA 94720, USA}

\author{E.~J.~Baxter~\orcidlink{0000-0002-6836-3196}}\affiliation{Institute for Astronomy, University of Hawai'i, 2680 Woodlawn Drive, Honolulu, HI 96822, USA}

\author{W.~d'Assignies~\orcidlink{0000-0002-9719-1717}}\affiliation{Institut de F\'{\i}sica d'Altes Energies (IFAE), The Barcelona Institute of Science and Technology, Campus UAB, 08193 Bellaterra (Barcelona) Spain}

\author{T.~Karwal~\orcidlink{0000-0002-1384-9949}}\affiliation{Department of Physics and Astronomy, University of Pennsylvania, Philadelphia, PA 19104, USA}

\author{A.~Amon~\orcidlink{0000-0002-6445-0559}}\affiliation{Department of Astrophysical Sciences, Princeton University, Peyton Hall, Princeton, NJ 08544, USA}

\author{D.~Anbajagane~\orcidlink{0000-0003-3312-909X}}\affiliation{Kavli Institute for Cosmological Physics, University of Chicago, Chicago, IL 60637, USA}

\author{F.~Andrade-Oliveira~\orcidlink{0000-0003-0171-6900}}\affiliation{Physik-Institut, University of Zürich, Winterthurerstrasse 190, CH-8057 Zürich, Switzerland}

\author{M.~R.~Becker~\orcidlink{0000-0001-7774-2246}}\affiliation{Argonne National Laboratory, 9700 South Cass Avenue, Lemont, IL 60439, USA}

\author{J.~Blazek~\orcidlink{0000-0002-4687-4657}}\affiliation{Department of Physics, Northeastern University, Boston, MA 02115, USA}

\author{A.~Carnero~Rosell~\orcidlink{0000-0003-3044-5150}}\affiliation{Instituto de Astrofisica de Canarias, E-38205 La Laguna, Tenerife, Spain}
\affiliation{Laborat\'orio Interinstitucional de e-Astronomia - LIneA, Av. Pastor Martin Luther King Jr, 126 Del Castilho, Nova Am\'erica Offices, Torre 3000/sala 817 CEP: 20765-000, Brazil}
\affiliation{Universidad de La Laguna, Dpto. Astrofísica, E-38206 La Laguna, Tenerife, Spain}

\author{C.~Chang~\orcidlink{0000-0002-7887-0896}}\affiliation{Department of Astronomy and Astrophysics, University of Chicago, Chicago, IL 60637, USA}
\affiliation{Kavli Institute for Cosmological Physics, University of Chicago, Chicago, IL 60637, USA}

\author{J.~M.~Coloma-Nadal~\orcidlink{0009-0003-0538-4349}}\affiliation{Institute of Space Sciences (ICE, CSIC),  Campus UAB, Carrer de Can Magrans, s/n,  08193 Barcelona, Spain}

\author{M.~Crocce~\orcidlink{0000-0002-9745-6228}}\affiliation{Institut d'Estudis Espacials de Catalunya (IEEC), 08034 Barcelona, Spain}
\affiliation{Institute of Space Sciences (ICE, CSIC),  Campus UAB, Carrer de Can Magrans, s/n,  08193 Barcelona, Spain}

\author{J.~DeRose~\orcidlink{0000-0002-0728-0960}}\affiliation{Lawrence Berkeley National Laboratory, 1 Cyclotron Road, Berkeley, CA 94720, USA}

\author{S.~Dodelson~\orcidlink{0000-0002-8446-3859}}\affiliation{Department of Astronomy and Astrophysics, University of Chicago, Chicago, IL 60637, USA}
\affiliation{Fermi National Accelerator Laboratory, P. O. Box 500, Batavia, IL 60510, USA}
\affiliation{Kavli Institute for Cosmological Physics, University of Chicago, Chicago, IL 60637, USA}

\author{G.~Giannini~\orcidlink{0000-0002-3730-1750}}\affiliation{Institute of Space Sciences (ICE, CSIC),  Campus UAB, Carrer de Can Magrans, s/n,  08193 Barcelona, Spain}
\affiliation{Kavli Institute for Cosmological Physics, University of Chicago, Chicago, IL 60637, USA}

\author{E.~Krause~\orcidlink{0000-0001-8356-2014}}\affiliation{Department of Physics, University of Arizona, Tucson, AZ 85721, USA}

\author{E.~Legnani~\orcidlink{0000-0001-7079-3796}}\affiliation{Institut de F\'{\i}sica d'Altes Energies (IFAE), The Barcelona Institute of Science and Technology, Campus UAB, 08193 Bellaterra (Barcelona) Spain}

\author{N.~MacCrann~\orcidlink{0000-0002-8998-3909}}\affiliation{Department of Applied Mathematics and Theoretical Physics, University of Cambridge, Cambridge CB3 0WA, UK}

\author{J. Mena-Fern{\'a}ndez~\orcidlink{0000-0001-9497-7266}}\affiliation{Aix Marseille Univ, CNRS/IN2P3, CPPM, Marseille, France}
\affiliation{Universit\'e Grenoble Alpes, CNRS, LPSC-IN2P3, 38000 Grenoble, France}

\author{A.~Porredon~\orcidlink{0000-0002-2762-2024}}\affiliation{Centro de Investigaciones Energ\'eticas, Medioambientales y Tecnol\'ogicas (CIEMAT), Madrid, Spain}
\affiliation{Ruhr University Bochum, Faculty of Physics and Astronomy, Astronomical Institute, German Centre for Cosmological Lensing, 44780 Bochum, Germany}

\author{J.~Prat~\orcidlink{0000-0002-5933-5150}}\affiliation{Departament de F\'isica, EEBE, Universitat Polit\`ecnica de Catalunya, c/ Eduard Maristany 16, E-08019 Barcelona, Spain}
\affiliation{University of Copenhagen, Dark Cosmology Centre, Juliane Maries Vej 30, 2100 Copenhagen O, Denmark}

\author{C.~Preston~\orcidlink{0000-0002-8113-6916}}\affiliation{Institute of Astronomy, University of Cambridge, Madingley Road, Cambridge CB3 0HA, UK}

\author{M.~Rodriguez-Monroy~\orcidlink{0000-0001-6163-1058}}\affiliation{Ruhr University Bochum, Faculty of Physics and Astronomy, Astronomical Institute, German Centre for Cosmological Lensing, 44780 Bochum, Germany}
\affiliation{Instituto de F\'isica Te\'orica UAM/CSIC, Universidad Aut\'onoma de Madrid, 28049 Madrid, Spain}
\affiliation{Laboratoire de physique des 2 infinis Irène Joliot-Curie, CNRS Universit\'e Paris-Saclay, Bât. 100, F-91405 Orsay Cedex, France}

\author{P.~Rogozenski~\orcidlink{0000-0002-1408-6904}}\affiliation{McWilliams Center for Cosmology and Astrophysics, Department of Physics, Carnegie Mellon University, Pittsburgh, PA 15213, USA}

\author{R.~Rosenfeld}
\affiliation{ICTP South American Institute for Fundamental Research\\ Instituto de F\'{\i}sica Te\'orica, Universidade Estadual Paulista, S\~ao Paulo, Brazil}
\affiliation{Laborat\'orio Interinstitucional de e-Astronomia - LIneA, Av. Pastor Martin Luther King Jr, 126 Del Castilho, Nova Am\'erica Offices, Torre 3000/sala 817 CEP: 20765-000, Brazil}

\author{S.~Samuroff~\orcidlink{0000-0001-7147-8843}}\affiliation{Department of Physics, Northeastern University, Boston, MA 02115, USA}
\affiliation{Institut de F\'{\i}sica d'Altes Energies (IFAE), The Barcelona Institute of Science and Technology, Campus UAB, 08193 Bellaterra (Barcelona) Spain}

\author{E.~Sanchez~\orcidlink{0000-0002-9646-8198}}\affiliation{Centro de Investigaciones Energ\'eticas, Medioambientales y Tecnol\'ogicas (CIEMAT), Madrid, Spain}

\author{I.~Sevilla-Noarbe~\orcidlink{0000-0002-1831-1953}}\affiliation{Centro de Investigaciones Energ\'eticas, Medioambientales y Tecnol\'ogicas (CIEMAT), Madrid, Spain}

\author{M.~Soares-Santos~\orcidlink{0000-0001-6082-8529}}\affiliation{Physik-Institut, University of Zürich, Winterthurerstrasse 190, CH-8057 Zürich, Switzerland}

\author{C.~To~\orcidlink{0000-0001-7836-2261}}\affiliation{Department of Astronomy and Astrophysics, University of Chicago, Chicago, IL 60637, USA}

\author{M.~A.~Troxel~\orcidlink{0000-0002-5622-5212}}\affiliation{Department of Physics, Duke University Durham, NC 27708, USA}

\author{M.~Yamamoto~\orcidlink{0000-0003-1585-997X}}\affiliation{Department of Astrophysical Sciences, Princeton University, Peyton Hall, Princeton, NJ 08544, USA}
\affiliation{Department of Physics, Duke University Durham, NC 27708, USA}

\author{B.~Yin~\orcidlink{0000-0001-7005-8820}}\affiliation{Department of Physics, Duke University Durham, NC 27708, USA}

\author{J.~Zuntz~\orcidlink{0000-0001-9789-9646}}\affiliation{Institute for Astronomy, University of Edinburgh, Edinburgh EH9 3HJ, UK}

\author{S.~Allam~\orcidlink{0000-0002-7069-7857}}\affiliation{Fermi National Accelerator Laboratory, P. O. Box 500, Batavia, IL 60510, USA}

\author{M.~Aguena~\orcidlink{0000-0001-5679-6747}}\affiliation{INAF-Osservatorio Astronomico di Trieste, via G. B. Tiepolo 11, I-34143 Trieste, Italy}
\affiliation{Laborat\'orio Interinstitucional de e-Astronomia - LIneA, Av. Pastor Martin Luther King Jr, 126 Del Castilho, Nova Am\'erica Offices, Torre 3000/sala 817 CEP: 20765-000, Brazil}

\author{D.~Bacon}
\affiliation{Institute of Cosmology and Gravitation, University of Portsmouth, Portsmouth, PO1 3FX, UK}

\author{D.~Brooks~\orcidlink{0000-0002-8458-5047}}\affiliation{Department of Physics \& Astronomy, University College London, Gower Street, London, WC1E 6BT, UK}

\author{R.~Camilleri}
\affiliation{School of Mathematics and Physics, University of Queensland,  Brisbane, QLD 4072, Australia}

\author{R.~Cawthon~\orcidlink{0000-0003-2965-6786}}\affiliation{Oxford College of Emory University, Oxford, GA 30054, USA}

\author{L.~N.~da Costa~\orcidlink{0000-0002-7731-277X}}\affiliation{Laborat\'orio Interinstitucional de e-Astronomia - LIneA, Av. Pastor Martin Luther King Jr, 126 Del Castilho, Nova Am\'erica Offices, Torre 3000/sala 817 CEP: 20765-000, Brazil}

\author{S.~Desai~\orcidlink{0000-0002-0466-3288}}\affiliation{Department of Physics, IIT Hyderabad, Kandi, Telangana 502285, India}

\author{P.~Doel~\orcidlink{0000-0002-6397-4457}}\affiliation{Department of Physics \& Astronomy, University College London, Gower Street, London, WC1E 6BT, UK}

\author{K.~Glazebrook}
\affiliation{Centre for Astrophysics \& Supercomputing, Swinburne University of Technology, Victoria 3122, Australia}

\author{K.~Honscheid~\orcidlink{0000-0002-6550-2023}}\affiliation{Center for Cosmology and Astro-Particle Physics, The Ohio State University, Columbus, OH 43210, USA}
\affiliation{Department of Physics, The Ohio State University, Columbus, OH 43210, USA}

\author{D.~Huterer~\orcidlink{0000-0001-6558-0112}}\affiliation{Department of Physics, University of Michigan, Ann Arbor, MI 48109, USA}

\author{K.~Kuehn~\orcidlink{0000-0003-0120-0808}}\affiliation{Australian Astronomical Optics, Macquarie University, North Ryde, NSW 2113, Australia}
\affiliation{Lowell Observatory, 1400 Mars Hill Rd, Flagstaff, AZ 86001, USA}

\author{J.~L.~Marshall~\orcidlink{0000-0003-0710-9474}}\affiliation{George P. and Cynthia Woods Mitchell Institute for Fundamental Physics and Astronomy, and Department of Physics and Astronomy, Texas A\&M University, College Station, TX 77843,  USA}

\author{R.~Miquel~\orcidlink{0000-0002-6610-4836}}\affiliation{Instituci\'o Catalana de Recerca i Estudis Avan\c{c}ats, E-08010 Barcelona, Spain}
\affiliation{Institut de F\'{\i}sica d'Altes Energies (IFAE), The Barcelona Institute of Science and Technology, Campus UAB, 08193 Bellaterra (Barcelona) Spain}

\author{M.~Paterno}
\affiliation{Fermi National Accelerator Laboratory, P. O. Box 500, Batavia, IL 60510, USA}

\author{A.~A.~Plazas~Malag\'on~\orcidlink{0000-0002-2598-0514}}\affiliation{Kavli Institute for Particle Astrophysics \& Cosmology, P. O. Box 2450, Stanford University, Stanford, CA 94305, USA}
\affiliation{SLAC National Accelerator Laboratory, Menlo Park, CA 94025, USA}

\author{M.~Smith~\orcidlink{0000-0002-3321-1432}}\affiliation{Physics Department, Lancaster University, Lancaster, LA1 4YB, UK}

\author{E.~Suchyta~\orcidlink{0000-0002-7047-9358}}\affiliation{Computer Science and Mathematics Division, Oak Ridge National Laboratory, Oak Ridge, TN 37831}

\author{D.~L.~Tucker~\orcidlink{0000-0001-7211-5729}}\affiliation{Fermi National Accelerator Laboratory, P. O. Box 500, Batavia, IL 60510, USA}

\author{V.~Vikram}
\affiliation{Central University of Kerala, Kasaragod, Kerala, India,  671325}

\author{J.~Weller~\orcidlink{0000-0002-8282-2010}}\affiliation{Max Planck Institute for Extraterrestrial Physics, Giessenbachstrasse, 85748 Garching, Germany}
\affiliation{Universit\"ats-Sternwarte, Fakult\"at f\"ur Physik, Ludwig-Maximilians Universit\"at M\"unchen, Scheinerstr. 1, 81679 M\"unchen, Germany}

\collaboration{DES Collaboration}

\begin{abstract}
We present a fast and numerically stable framework for conducting posterior predictive distribution (PPD) tests of internal consistency in correlated cosmological probes.  
Our method, developed as a validation check for the Dark Energy Survey (DES) Year 6 combined analysis of weak lensing and galaxy clustering, employs a Gaussian-mixture model to efficiently approximate the high-dimensional PPD and introduces  a scalar consistency metric, $\Delta_{\rm PPD}$, defined as the fraction of the PPD at lower probability density than the observed data. 
This method avoids challenges related to calibration and numerical stability that impacted PPD methods previously used in DES Year~3. 
After demonstrating performance with a toy model, we confirm that when applied to DES Year~3 measurements, $\Delta_{\rm PPD}$ identifies the same inter-probe tensions as published, calibrated PPD results while requiring only minutes per test. We further validate the method on several noise realizations of simulated DES Y6 analyses. These studies demonstrate that $\Delta_{\rm PPD}$ responds coherently to injected inconsistency, while additionally highlighting two effects which can potentially complicate the interpretation of PPD tests more generally in the context of increasingly complex models. First, internal tensions that can be absorbed as shifts in nuisance parameters become more difficult to detect. Second,  poorly constrained directions in parameter space  can induce projection effects in PPDs for disjoint sets of observables, producing reports of tension even for consistent data. 
Given this, we conclude that $\Delta_{\rm PPD}$ provides a practical PPD-based diagnostic for multiprobe unblinding, though it should be interpreted with care alongside other validation checks. 
This framework has already played an important role in DES analyses and is well suited for application to forthcoming Stage-IV surveys such as LSST, Euclid, and Roman, where fast, robust internal consistency tests will be essential.
\end{abstract}


\maketitle

\section{Introduction}

While the growing volume of cosmological data represents great promise for furthering our understanding of the Universe, it also necessitates advances in methodology to ensure analyses are accurate and robust.
A crucial component of this effort is the development and characterization of internal consistency tests. This is because many  studies, including recent  joint analyses of galaxy clustering and weak lensing~\citep*{y3-3x2ptkp,2007.15632,2304.00704,2304.00705},  rely on the combined analysis of multiple correlated observables to separate signals of cosmological physics from astrophysical and measurement uncertainties. Internal consistency tests can include goodness-of-fit assessments, comparisons of parameter constraints derived from data subsets, and model comparison~\citep{1703.05959,1809.07333,1809.01406,2019PhRvD..99d3506R,1910.07820,2020PhRvD.101j3527R,1909.10991,2007.08496,2105.03324,2407.15478,2503.19442}. Tests that can be carried out without revealing cosmology results play a vital role in blind analysis methods~\citep*{y3-blinding} when used as pre-unblinding validation checks.  In this paper we present a novel approach to conducting internal consistency tests based on posterior predictive distributions (PPD), developed as part of the Dark Energy Survey \citep[DES,][]{1601.00329} Year 6 (Y6) combined analysis of galaxy clustering and weak lensing~\citep*{y6-3x2pt}. Our focus is on the "$3\times2$pt" analysis, which analyzes three two-point correlation functions---cosmic shear, galaxy-galaxy lensing (GGL), and galaxy clustering---in  a unified framework, making it a cornerstone of modern cosmological analyses.

The PPD is a Bayesian construction~\citep{2014bda..book.....G} for quantifying the consistency of observables conditioned on an assumed model. It is formally defined as the convolution of a posterior distribution in parameter space with a likelihood function, yielding a probability distribution in data space. 
Given posterior samples (e.g., from MCMC), the PPD can be readily sampled by drawing from the likelihood evaluated at each posterior sample, without needing to compute this convolution explicitly.
Comparing new data to that expected distribution allows us to assess consistency between different measurements. 
In practice, PPD tests involve applying a summary function to compress the high-dimensional data-space distribution into  consistency metrics which can be practically employed as part of analysis validation checks. The underlying rationale is that if the model and data are consistent, the summary of the observed data should lie within the bulk of the PPD summary distribution, and measurements that are extreme relative to that distribution could indicate problems with the analysis. This is intuitively analogous to reduced-$\chi^2$ goodness-of-fit tests, where a too-high or too-low minimum $\chi^2$ indicates, respectively, a model's inability to fit the data or underestimated error bars.
PPD methods have an advantage over reduced $\chi^2$ tests in that they do not rely on point-estimates of the best-fit parameters, which tend to be noisy and difficult to estimate in high-dimensional parameter spaces.  This analogy is not incidental: although the predictive distribution is Bayesian, comparing the observed data to hypothetical replications is a sampling-theory calculation. The PPD thus provides a frequentist form of model criticism built on Bayesian estimation~\citep{Box1980,Rubin1984}.

PPD tests have previously been applied in the DES Year~3 clustering and lensing analyses~\citep[DES~Y3,][]{y3-cosmicshear1,y3-3x2ptkp,y3-2x2ptaltlenssompz,y3-2x2ptmagnification} and in a recent LSST modified-gravity forecast to quantify agreement with general relativity~\citep{2511.19194}. The Kilo-Degree Survey (KiDS) lensing analyses~\citep{2503.19442,2007.15633} employ a related technique, the translated posterior distribution~\citep{1809.01406}.  
Beyond cosmology, PPD methods find applications in diverse quantitative fields, including ecology~\citep{10.1002/ece3.993,10.1002/ecm.1314}, attesting to their generality as a model-checking framework. 

Translating high dimensional PPDs into interpretable consistency metrics, and in particular calibrating them to obtain formal $p$-values, remains an open problem in statistics. Here a $p$-value refers to the probability, assuming a model is correct, that the observables (or a summary function thereof) will be at least as extreme as the measured data.   
Existing approaches in statistics often assume that independent and identically distributed data realizations are available~\citep{1908.00882,2203.15897,2112.03333}, and are therefore ill-suited for large-scale structure cosmological analyses where only a single data realization exists. In previous large-scale structure analyses, including the  DES Y3  combined analysis of galaxy clustering and weak lensing, a commonly used summary function is $\chi^2$ difference between model predictions and either the observed data or realizations of simulated observables drawn from the PPD. Compressing this into a consistency metric remains difficult, as this comparison requires analyzing many synthetic data realizations at each of many locations in parameter space, introducing challenges related to computational cost and numerical stability. This process additionally requires   calibration via simulated analyses to obtain a $p$-value, as uncalibrated PPD $\chi^2$ metric can be severely biased towards reporting inconsistency~\citep{1802.05257,2203.07128}. While this DES Y3 PPD methodology was effective enough to serve as a key component of the DES Y3 analysis validation, the associated computational  challenges made the process labor-intensive and thus a bottleneck during unblinding. 
   
We propose two innovations to address these challenges: a numerically stable approximation of the PPD using a Gaussian mixture model, and a consistency metric, denoted $\Delta_{\rm PPD}$, that leverages this approximation to efficiently quantify agreement between data and model in an interpretable way even without formal calibration. A key feature of both developments is their streamlined implementation, designed so that the amount of work needed to conduct PPD tests is comparable to other MCMC chain post-processing operations routinely performed in cosmology studies.  After introducing this PPD methodology, we present three sets of tests used to characterize its performance. A toy model study serves as an initial test of the Gaussian mixture PPD estimate and $\Delta_{\rm PPD}$ metric, and builds intuition for what this metric captures. Next, we apply it to DES Y3 galaxy-clustering and weak-lensing measurements, to assess the agreement between our new  $\Delta_{\rm PPD}$ metric and the calibrated $p$-value previously employed in that analysis. Finally we study the behavior of our PPD consistency metric when applied to simulated DES Y6 galaxy-clustering and weak-lensing analyses, with and without injected inconsistencies. 

While in general our updated PPD framework performs well---particularly regarding improved ease of use and numerical stability compared to previous methods---two findings from our simulated DES Y6 tests merit closer examination. These will be discussed in more detail below, but we highlight them here as important considerations when incorporating PPD tests into an analysis validation strategy. 
First, we note that the PPD consistency metric can be less sensitive to an injected inconsistency than one might naively expect. If the model is sufficiently flexible, data contamination may shift nuisance parameter posteriors rather than $\Delta_{\rm PPD}$ itself. 
Second, for consistency tests between two different observables, PPD tests  can be subject to projection effects when the test and conditioning observables are sensitive to different degenerate combinations of model parameters. Projection effects, studied in the context of their impact on marginalized parameter constraints in e.~g.~\citet*{y6-methods}, \citet{2411.12021,2408.07700,2409.09101,2506.12004}, can skew PPD metrics including $\Delta_{\rm PPD}$ towards values indicating inconsistency even for consistent input data.  
This can be accounted for via calibration studies, or more simply --- and as was done in practice for DES Y6 --- by using simulated analyses to guide how PPD tests should be used during pre-unblinding checks. Taken together, these results support the idea that this PPD framework is a useful diagnostic tool for complex cosmological analyses, but one that should be used with care and in conjunction with complementary validation tests.

The remainder of this paper is organized as follows. In \cref{sec:formalism}, we review posterior predictive distributions, describe our Gaussian–mixture approximation to the PPD, and introduce our new density–based PPD metric, $\Delta_{\rm PPD}$. In \cref{sec:y3}, we apply the method to DES~Y3 data and compare its behavior to the published, calibrated PPD tests used in that analysis. \Cref{sec:y6} presents our validation of the method on simulated DES~Y6 $3\times2$pt data, including tests with and without an injected $X_{\rm lens}$ inconsistency, and discusses effects that can complicate the interpretation of $\Delta_{\rm PPD}$: absorption of tensions by nuisance parameters, and projection effects in disjoint PPD tests. We summarize our findings and discuss the role of $\Delta_{\rm PPD}$ within broader consistency tests in \cref{sec:summary-discussion}.

\section{Formalism}
\label{sec:formalism}

We consider two sets of measurements from one or several experiments, $\vd$ and $\vdp$. At this stage, $\vd$ and $\vdp$ may correspond to (i) independent realizations of the same measurement, or (ii) disjoint, but potentially correlated, measurements. The vector $\bth$ represents  parameters of the statistical model for those observables. We use $\mathcal{N}(\vd; \bm{\mu}_\vd,\mathbf{C})$ to denote a multivariate Gaussian distribution on $\vd$ with mean $\bm{\mu}_\vd$ and  covariance $\mathbf{C}$. 

\subsection{Definitions}

The posterior predictive distribution (PPD) is the distribution of future realizations of the measurement $\vdp$, conditioned on the observation of a previous realization of $\vd$, labeled $\vdobs$. It is given by
\begin{equation}
    \Prob{\vdp | \vdobs} = \int \Prob{\vdp | \vdobs, \bth} \Prob{\bth | \vdobs} \dd{\bth}.
    \label{eq:def_ppd}
\end{equation}
Here $\Prob{\bth | \vdobs}$ is the posterior distribution inferred from $\vdobs$, while $\Prob{\vdp | \vdobs, \bth}$ is the conditional distribution of $\vdp$ given model parameters $\bth$ and the observed data $\vdobs$. In other words, the PPD is obtained by marginalizing the conditional distribution of $\vdp$ over the posterior uncertainty on $\bth$ given $\vdobs$. 
As a result, the PPD is generally broader than the likelihood based on a point estimate of $\bth$ (e.g., the best fit), since it incorporates both parameter uncertainty and the intrinsic scatter described by $\Prob{\vdp | \vdobs, \bth}$. 
While this integral is intractable in general, it can be estimated using a Monte-Carlo sampling method that we propose in the following sections.

To build intuition for what information the PPD captures, it is instructive to consider the analytically tractable case of a linear Gaussian model, where $\bm{\mu}_\vd(\bth)=\mathbf{M}_{\vd}\bth$ and $\bm{\mu}_\vdp(\bth)=\mathbf{M}_{\vdp}\bth$. Assuming a Gaussian likelihood with covariance $\mathbf{C}_{\vd}$ and a Gaussian prior
${\Prob{\bth} = \mathcal{N}(\bth; \bth_0, \mathbf{\Sigma}_0)}$,
the posterior is also Gaussian. The posterior mean $\bm{\theta}_p$ and covariance $\mathbf{\Sigma}_p $ are given by
\begin{align}
    \bm{\theta}_p &=\mathbf{\Sigma}_p \qty(\mathbf{\Sigma}_0^{-1}\bth_0 + \mathbf{M}_{\vd}^T \mathbf{C}_{\vd}^{-1}\vdobs), \\
    \mathbf{\Sigma}_p &= \qty(\mathbf{\Sigma}_0^{-1} + \mathbf{M}_{\vd}^T \mathbf{C}_{\vd}^{-1} \mathbf{M}_{\vd})^{-1},
\end{align}
where the mean coincides with the posterior maximum~\citep[see, e.g.,][]{2019PhRvD..99d3506R}.
If we also assume that $\vd$ and $\vdp$ are disjoint and uncorrelated, the PPD is then also Gaussian with mean $\vb{y}_{\rm PPD}$ and covariance $\mathbf{C}_{\rm PPD}$ given by
\begin{align}
    \vb{y}_{\rm PPD} &= \mathbf{M}_{\vdp}\bm{\theta}_p,\\
    \mathbf{C}_{\rm PPD} &= \mathbf{C}_{\vdp}+\mathbf{M}_{\vdp}\mathbf{\Sigma}_p^{-1}\mathbf{M}_{\vdp}^T\label{eq:ppd_cov}.
\end{align}
Here the two covariance terms capture the contributions from data noise and parameter uncertainty, respectively. This is in contrast to a $\chi^2$ goodness-of-fit test, which conditions on a point estimate of $\bth$ and therefore accounts only for data noise---a distinction that proves important when the posterior is broad, as the PPD will then be wider than the likelihood alone (a visual example is provided in \cref{sec:toy}).

A PPD consistency test is typically based on a scalar-valued function $T(\vdp)$, and in particular, on the comparison of its observed value $T(\vdpobs)$ to the distribution of $T(\vdp)$ from PPD samples $\vdp\sim\Prob{\vdp | \vdobs}$. More generally, one may define a parameter-dependent summary $T(\vdp,\bth)$ where, at each posterior sample $\bth \sim \Prob{\bth | \vdobs}$, $T(\vdpobs,\bth)$ is compared to the distribution of $T(\vdp,\bth)$ sampled at $\vdp\sim\Prob{\vdp | \vdobs, \bth}$. The fraction of PPD samples with a summary function lower (or higher) than that computed on the data defines a PPD-based metric in the range $[0,1]$: values too close to 0 indicate inconsistency while values too close to 1 indicate either  over-estimated uncertainty or underestimated cross-correlations. As noted in the introduction, this is analogous to how we interpret too-low or too-high values for the minimum $\chi^2$ per degree of freedom in point-estimate goodness-of-fit tests.

If the likelihood $\mathcal{L}\left(\vdp | \bth\right)\equiv\Prob{\vdp | \bth}$ is Gaussian, then a natural choice for $T(\vdp,\bth)$ is the $\chi^2$ difference between $\vdp$ and $\bm{\mu}_\vdp(\bth)$. As we shall see, this parameter-dependent metric has drawbacks, especially when considering consistency tests of disjoint data sets, as explored in~\citet{2203.07128}. We instead propose to use the density of the PPD itself, $\Prob{\vdp | \vdobs}$, which will become tractable even in high-dimensions for the cosmological analyses we target, given few additional assumptions.

\subsection{Approximating the PPD for Gaussian likelihoods}

For $3\times2$pt analyses we typically assume a Gaussian likelihood with a parameter-independent covariance.\footnote{While the covariance of $3\times 2$pt measurements does depend on cosmology, for typical analyses it is simply evaluated at a fixed set of fiducial parameter values.  \citet*{y3-covariances} showed that doing so has negligible impact on cosmological inference. The assumption of Gaussian likelihood for weak lensing was studied in, e.g.~\citet{2022PhRvD.105l3527H}.} This assumption allows us to introduce an efficient approximation for the PPD using a Gaussian mixture model.

For a set of posterior samples  $\qty{\bth_i}$ with weights $\qty{w_i}$ such that ${\sum_i w_i = 1}$, a Monte-Carlo estimate of the integral in \cref{eq:def_ppd} may be obtained by writing the PPD as a mixture of conditional likelihood distributions evaluated at the posterior samples, such that
\begin{equation}
    \Prob{\vdp | \vdobs} \approx \sum_i w_i \Prob{\vdp | \vdobs, \bth_i}.
    \label{eq:mc_ppd}
\end{equation}
We distinguish two cases:
\begin{enumerate}
    \item \textit{Goodness-of-fit tests}. If $\vd$ and $\vdp$ refer to independent measurement realisations where $\vdp$ represents either the same observables as $\vd$ or a subset of them,  
    then $\Prob{\vdp | \vdobs, \bth}$ reduces to the likelihood $\mathcal{L}\left(\vdp | \bth\right)$. If the likelihood is a multivariate Gaussian around a mean $\bm{\mu}_\vdp(\bth)$ with parameter-independent covariance $\mathbf{C}_{\vdp\vdp}$, then \cref{eq:mc_ppd} can be written
    \begin{equation}\label{eq:ppd_gauss_mix_gof}
        \Prob{\vdp | \vdobs} \approx \sum_i w_i \,\mathcal{N}(\vdp; \bm{\mu}_\vdp(\bth_i),\mathbf{C}_{\vdp\vdp}).
    \end{equation}
    This is a mixture of multivariate Gaussian distributions with matching covariances but different means, with one Gaussian  for each posterior sample $\bth_i$. 
    \item \textit{Disjoint internal consistency tests}. If $\vd$ and $\vdp$ refer to disjoint components of a set of correlated measurements, then it may be challenging to analytically express the conditional distribution $\Prob{\vdp | \vdobs, \bth}$ in the general case. However, if the likelihood of the full measurement of both $\vd$ and $\vdp$ is a multivariate Gaussian, then so are the conditional distributions.
    For disjoint data vector subsets $\vd$ and $\vdp$, we write
    \begin{equation}
        \Prob{{\mqty(\vd \\ \vdp)} | \bth} = \mathcal{N} \qty( 
        \mqty(\vd \\ \vdp ); \mqty(\mu_\vd(\bth) \\ \mu_\vdp(\bth)), \mqty(\mathbf{C}_{\vd\vd} & \mathbf{C}_{\vd\vdp} \\ \mathbf{C}_{\vdp\vd} & \mathbf{C}_{\vdp\vdp})),
    \end{equation}
    where $\mathbf{C}_{\vd\vdp}$ represents the cross-covariance matrix between $\vd$ and $\vdp$. It can be shown that $\Prob{\vdp | \vd, \bth}$, the distribution of $\vdp$ conditioned on $\vd$ and $\bth$,  is then also a multivariate Gaussian with mean ${\mu_{\vdp \mid \vd}(\bth)}$ and covariance ${\mathbf{C}_{\vdp \mid \vd}}$ given by
    \begin{align}
        \mu_{\vdp \mid \vd}(\bth) & = \mu_\vdp(\bth) + \mathbf{C}_{\vdp\vd} \mathbf{C}_{\vd\vd}^{-1} (\vd - \mu_\vd(\bth)), \\
        \mathbf{C}_{\vdp \mid \vd} & = \mathbf{C}_{\vdp\vdp} - \mathbf{C}_{\vdp\vd}\mathbf{C}_{\vd\vd}^{-1} \mathbf{C}_{\vd\vdp}.
    \end{align}
    Given a measurement of $\vdobs$ and  associated posterior samples $\qty{\bth_i}$  with weights $\qty{w_i}$, we thus may approximate \cref{eq:mc_ppd} as a mixture of multivariate Gaussian distributions
    \begin{align}
    \label{eq:ppd_gauss_mix_disjoint}
        \Prob{\vdp | \vdobs} \approx \sum_i w_i \mathcal{N} \qty( \vdp; \mu_{\vdp \mid \vd}(\bth_i),
        \mathbf{C}_{\vdp \mid \vd} ).
    \end{align}
    Again, this corresponds to a mixture of multivariate Gaussians with varying means. Note that if the covariance is assumed to be independent of model parameters, the covariance matrix for the conditional distribution is also independent of model parameters. 
\end{enumerate}

In both above cases, the posterior predictive distribution is approximated by a mixture of multivariate Gaussians. We can then sample from that approximate distribution and compute the probability density function for a given data realization (observed or simulated). This enables us to define and efficiently compute a new, density-based PPD metric, which will be described in the following section.

\subsection{PPD consistency metric}
\label{sec:metric}

In practice, to use the PPD as a tool for validating analyses, we must compress this probability distribution defined over high-dimensional data space into an interpretable summary statistic. To this aim, we define a consistency metric, $\Delta_{\rm PPD}$, given by the probability that a data realization drawn from the PPD  has lower probability than the observed data. In other words, this quantity is the fraction of the posterior predictive distribution at lower density than the iso-contour corresponding to the  probability of the observed data realization $\vdpobs$.
This is formally given by
\begin{equation}
    \Delta_{\rm PPD}\qty(\vdpobs | \vdobs) \equiv \int_{\Prob{\vdp | \vdobs} < \Prob{\vdpobs | \vdobs}} \Prob{\vdp | \vdobs} \dd{\vdp}.
    \label{eq:delta_ppd}
\end{equation}
A low value of $\Delta_{\rm PPD}$ indicates that the observed data $\vdpobs$ falls in a low-density region of the PPD relative to typical PPD samples, signaling inconsistency of the data with the model used to describe them. This is similar to the metric defined  to quantify tensions in parameter space for difference distributions in~\citet{2105.03324}, and corresponds to using the density of the PPD as summary function, \ie $T(\vdp)=\Prob{\vdp | \vdobs}$.

\Cref{eq:delta_ppd}  may be evaluated numerically to arbitrary precision by rejection sampling. To do so, we generate samples $\vdp\sim\Prob{\vdp | \vdobs}$ from the approximate Gaussian mixture and compute the corresponding probability density for each. By comparing these to the density evaluated at the observed data vector $\Prob{\vdpobs | \vdobs}$, we can find the fraction of samples with lower probability, which serves as an unbiased estimator of $\Delta_{\rm PPD}$. Note that for goodness-of-fit tests, the PPD is a distribution over hypothetical future data, but the test assesses the position of the actual observed data within this distribution. Here, the posterior estimate and PPD metric are both evaluated for the same realization, so with $\vdpobs=\vdobs$. In contrast, for disjoint tests, the PPD is conditioned on one observed dataset, $\vdobs$, and assesses the position of a separate observed dataset, $\vdpobs$.

The reliability of this estimator depends on the fidelity of the PPD approximation, which is tied to the accuracy of the posterior estimate. Given a posterior sample, the stability of the mixture model can be tested by bootstrapping posterior samples and recomputing the PPD and the $\Delta_{\rm PPD}$ metric. This procedure  yields stable results across all tests presented here. This robustness is expected, since the data covariance governs the effective width of both the likelihood and the posterior, ensuring that representative posterior samples map faithfully into data space through the PPD.

The metric $\Delta_{\rm PPD}$ has the desirable characteristic that for consistency tests over disjoint data subsets $\vd$ and $\vdp$, if $\vd$ is weakly informative about $\vdp$ (in the sense that  $\vdp$ depends on model parameters that are poorly constrained by $\vd$), the test should generally favor consistency. This is because we expect that for weak constraints,  the PPD $\Prob{\vdp | \vdobs}$ will be a broad distribution dominated by parameter uncertainty rather than data noise, and thus likely to have high probability density at $\vdpobs$, resulting in values of $\Delta_{\rm PPD}$ that indicate consistency. 
Note that this expectation may not hold in certain scenarios where $\vdp$  has greater sensitivity to parameters that are weakly constrained by $\vd$, as  projection effects can shift the PPD away from $\vdpobs$ and bias $\Delta_{\rm PPD}$ towards inconsistency, as will be discussed in \cref{sec:y3,sec:y6}.

It is also important to note that $\Delta_{\rm PPD}$ is not guaranteed to be uniformly distributed over $[0,1]$ under the null hypothesis of model-data consistency, and therefore does not constitute a $p$-value in a formal  sense. In \cref{app:pvalue}, we demonstrate this explicitly using the toy model presented in the next section.  
Deriving a formal $p$-value from $\Delta_{\rm PPD}$ would require characterizing its sampling distribution under data replication, which is not uniquely defined in a Bayesian setting. This was nevertheless done for the calibrated DES~Y3 PPD metric by repeatedly generating mock data  at best fit  and computing the PPD metric over those realizations~\citep{2203.07128}, as discussed in \cref{sec:DESY3method} below.\footnote{Alternative  approaches to calibrating PPD metrics could include analytical calculation, which is typically intractable, or leveraging independent and identically distributed data subsets, which is not feasible for large-scale structure analyses \citep[see, e.g.,][]{1908.00882,2203.15897,2112.03333}.}
In this work, we do not perform such a calibration for $\Delta_{\rm PPD}$. Instead, in \cref{sec:y3,sec:y6} we demonstrate on both real and simulated data that $\Delta_{\rm PPD}$ itself generally serves as a reliable diagnostic indicator of internal consistency---while also identifying certain regimes where its interpretation requires care.

\subsection{Toy model}
\label{sec:toy}

\begin{figure*}
    \centering
    \begin{minipage}[c]{.33\textwidth}
        \includegraphics[scale=0.65]{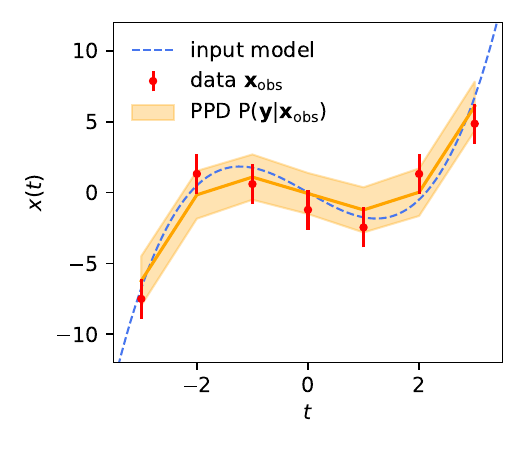}
    \end{minipage}%
    \begin{minipage}[c]{.33\textwidth}
        \includegraphics[scale=0.65]{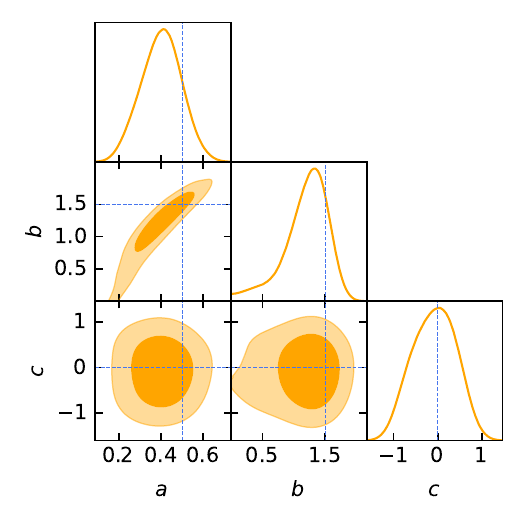}
    \end{minipage}%
    \begin{minipage}[c]{.33\textwidth}
        \includegraphics[scale=0.65]{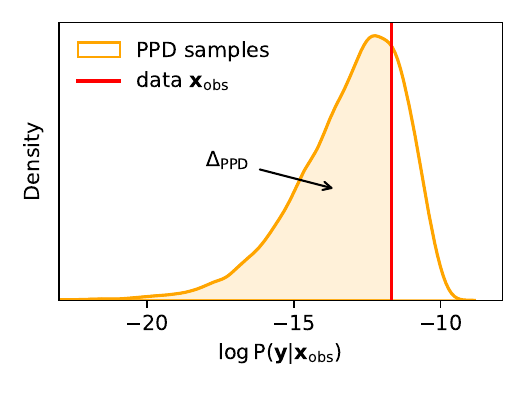}
    \end{minipage}
    \caption{
    Illustration of the toy model PPD computation for a goodness-of-fit test, where $\vd$ and $\vdp$ denote independent realizations of the same observables. Left: model at input parameters (blue), an observed realization $\vdobs$ sampled from the likelihood (red), and the PPD band $\Prob{\vdp | \vdobs}$ showing its mean $\pm 1 \sigma$ (orange). Middle: posterior distributions of $a$, $b$, and $c$ for this realization, with input parameters indicated by dashed lines. Right: histogram of log-density values for PPD samples (smoothed with a kernel density estimate) compared to the PPD log-density for the observed data (evaluated for $\vdp=\vdobs$), with $\Delta_{\rm PPD}$ corresponding to the shaded area. }
    \label{fig:toy-ppd}
\end{figure*}

To build intuition for the  $\Delta_{\rm PPD}$ statistic, we illustrate our PPD framework  with a concrete numerical example using a simple toy model.  Consider a data vector denoted by ${\vd = \qty(x_1, \dots, x_n)}$, corresponding to samples of a parameterized curve $x(t)$ evaluated at $n$ fixed times, $t_i$ with $i\in\{1,\dots,n\}$, and let ${\vt=\qty(t_1, \dots, t_n)}$. Let us assume the theoretical prediction is
\begin{equation}
    \bm{\mu}_{\vd}(\bth)(t) = a\,t^3 - b^2\,t + c^3,
\end{equation}
with parameters $\bth=(a,b,c)$. The likelihood then takes the Gaussian form
    ${\mathcal{L}(\vd \mid \bth) = \mathcal{N}\qty(\vd; \mu_{\vd}(\bth), \mathbf{C})}$,
where $\mathbf{C}$ is a fixed covariance matrix with mild correlations between adjacent points. For illustration we choose input values $a_0=0.5$, $b_0=1.5$, and $c_0=0$. The priors are uniform for $a$ and $b$ with ranges $[0,1]$ and $[0,3]$, respectively, while $c$ follows a unit Gaussian prior.

As illustrated in \cref{fig:toy-ppd}, we first evaluate the model at the input parameters (blue line) and draw a realization from the likelihood (red points with error bars), shown in the left panel. We then estimate the posterior distributions of $a$, $b$, and $c$ from this realization (middle panel) and use the resulting posterior samples to generate the PPD. The PPD is shown in orange in the left panel, with the line denoting the PPD mean and the shaded region showing its standard deviation. Finally, we sample from the PPD to compute a histogram of probability-density values and compare them with the observed data (right panel). The quantity $\Delta_{\rm PPD}$ indicates the fraction of PPD samples with probability density below that of the observed data. In this demonstration we find $\Delta_{\rm PPD}=0.61$, indicating internal consistency.

\subsection{Implementation}

The computation of $\Delta_{\rm PPD}$ can be practically separated into two stages. The first step is to  calculate   model predictions for the observables being tested at each posterior sample ($\{\bm{\mu}_\vdp(\bth_i)\}$ in \cref{eq:ppd_gauss_mix_gof} and \cref{eq:ppd_gauss_mix_disjoint}). This step relies on the same calculations, and thus the same modeling software,  used for parameter estimation. In principle, this can be done without additional computation if the predictions  are saved as derived parameters when running initial posterior estimation chains. Otherwise, 
one must parse an existing posterior estimation chain, then compute and save model predictions for each sample.
In practice for DES Y6 we do this using the \textsc{CosmoSIS} importance sampler~\citep{cosmosis2020}, which allows us to easily run our modeling pipeline for each sample in an initial chain, saving the theoretical prediction for the full $3\times2$pt data vector\footnote{One exception here is that we do not save the full $3\times2$pt data vector for cosmic shear chains because computing galaxy-clustering predictions requires parameters (e.g. galaxy bias) not sampled in the cosmic shear chains.}  using the software's \texttt{extra\_output} setting.\footnote{As an additional practical consideration, before running the importance sampler to save model predictions, we produce an equal-weight subsample of the initial chain, sized based on the effective number of samples in the initial chain. This is because the {Nautilus}~\citep{2024AAS...24422902L} sampler used to run most DES~Y6 chains saves millions of posterior samples per chain, though the effective sample size is $\sim \mathcal{O}(10^5)$. Working with a subsample thus makes computational time and file sizes more manageable.} The distribution of these model predictions gives the posterior for the associated observables which we must convolve with a distribution accounting for data noise to obtain the PPD.

The second stage of our calculation accounts for this by generating   a Gaussian mixture PPD from those model predictions.  This step is independent of the modeling pipeline (e.g., of \textsc{CosmoSIS}).  
We employ the \textsc{TensorFlow Probability} framework~\citep{1711.10604} to efficiently compute the approximate (log) density of the PPD from \cref{eq:ppd_gauss_mix_gof,eq:ppd_gauss_mix_disjoint}. Its implementation of mixture distributions leverages GPU vectorization and memory-efficient operations, allowing fast evaluation once posterior samples and corresponding means are available. With our implementation, computing $\Delta_{\rm PPD}$ for a given posterior typically takes only a few minutes.

The code for this implementation is publicly available at \href{https://github.com/des-science/y6-ppd-public/}{\texttt{https://github.com/des-science/y6-ppd-public/}}.  

\section{Application to DES~Y3 data}
\label{sec:y3}

PPD tests were used in DES~Y3 as part of the pre-unblinding validation of the joint analyses of galaxy clustering, galaxy--galaxy lensing (GGL), and cosmic shear presented in \citet*{y3-cosmicshear1,y3-3x2ptkp,y3-2x2ptaltlenssompz,y3-2x2ptmagnification}. These tests evaluated goodness-of-fit for the $1\times2$ (shear), $2\times2$ (clustering plus GGL), and $3\times2$ (all three) combinations, as well as internal consistency tests in which one two-point function is conditioned on the other two, all under a $\Lambda$CDM model. As an initial validation of $\Delta_{\rm PPD}$, we repeat these tests and compare our results to the published calibrated values.

Before presenting the comparison, we emphasize one subtlety that affects consistency tests between disjoint sets of observables in particular. These tests are sensitive to how informative the conditioning data are about the probe being tested, as mentioned in \cref{sec:metric}. 
As an example relevant to the studies presented below, both GGL and galaxy-clustering measurements are sensitive to galaxy bias parameters, but for GGL that dependence is approximately linear, while for galaxy clustering it is approximately quadratic.\footnote{These dependencies will be exact for a linear galaxy bias model if we neglect clustering contributions from redshift space distortions and magnification.}
That stronger sensitivity to galaxy bias means that the PPD of galaxy clustering given GGL plus shear will yield a broad distribution, resulting in a statistically weak test of internal consistency. 
This outcome is driven by the limited information that GGL provides on clustering biases.  In the following sections, we review how this effect impacted the DES~Y3 metric and imposed a calibration step, and demonstrate how our proposed metric, $\Delta_{\rm PPD}$, provides interpretable outputs without calibration.

\subsection{Calibration challenges of the DES Y3 PPD method}
\label{sec:DESY3method}

\begin{figure*}
    \centering
    \includegraphics[scale=0.65]{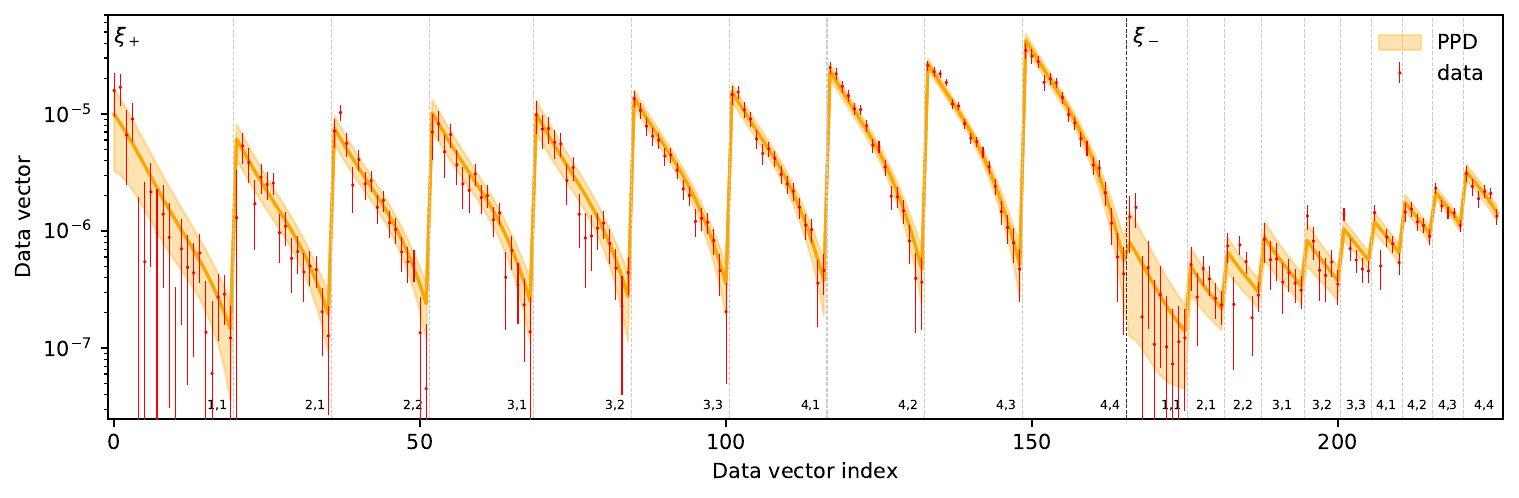}
    \caption{Visual representation of the goodness-of-fit test for DES~Y3 cosmic shear data ($\xi_+$ and $\xi_-$ are stacked) with standard scale cuts. Real data points are shown in red, while the PPD is represented by the orange band (analytic computation) representing the mean of the distribution, plus or minus one standard deviation. Grey dashed lines separate the measurements from different tomographic bin pairs, which are indicated at the bottom.}
    \label{fig:ppd_dv}
\end{figure*}

To provide context for comparison with our new PPD framework, here we briefly outline the DES~Y3 PPD methodology and the calibration challenges it presented, which are described in detail in \citet{2203.07128}. At each posterior sample $\bth_i$, a theory prediction $\bm{\mu}_\vdp(\bth_i)$ is computed and two $\chi^2$ values are evaluated: one quantifying the difference between $\bm{\mu}_\vdp(\bth_i)$   and the observed data $\vdpobs$, and one  comparing $\bm{\mu}_\vdp(\bth_i)$ and a simulated data realization $\vdp$ drawn from the PPD (in practice, drawn from the Gaussian likelihood for the observables evaluated at $\bth_i$). For each posterior sample, we can evaluate whether the theory-vs-PPD-draw $\chi^2$  is larger than the theory-vs-data $\chi^2$.  The consistency metric for DES~Y3 was defined as the fraction of posterior samples where that condition holds:  where there is better agreement between the data and model prediction  than between the model and random PPD draws. 
\footnote{In practice, several PPD samples are drawn at each point to reduce noise in the metric estimate.} While intuitive, this approach has a key limitation, in that it is systematically biased towards inconsistency for tests where the width of the PPD is dominated by parameter uncertainty rather than by the data covariance, because in that case the theory-vs-data $\chi^2$ is then systematically larger than the theory-vs-PPD $\chi^2$. 
Obtaining interpretable $p$-values therefore requires calibration. 

For DES~Y3 analyses, calibration was achieved by first evaluating this $\chi^2$-based PPD metric for many noise realizations for simulated data, drawn  around the best-fit model prediction. This step samples the expected distribution of the uncalibrated metric assuming the best-fit parameters. We can then compare that distribution to the PPD metric evaluated for the observed data. The fraction of simulated data analyses for which the $\chi^2$-based PPD metric is lower than for real data can then be interpreted as a calibrated $\tilde{p}$-value. Difficulties with this approach arose because running a full posterior analysis for each simulated realization was computationally prohibitive. In practice this calibration step was based on approximate posteriors for the simulated data obtained by importance sampling the observed data's posterior. That importance sampling step was numerically challenging and required careful fine-tuning to yield stable results. Although the published calibrated DES~Y3 $\tilde{p}$-values were verified to be accurate, these difficulties directly motivated the development of the new PPD methodology presented in this paper. Because our new Gaussian-mixture model approximation for the PPD and the associated $\Delta_{\rm PPD}$ metric  generally yield interpretable metrics without calibration, this lessens the need for the numerically unstable importance sampling step. 

\subsection{Results}

Like the DES~Y3 approach, our new PPD method enables intuitive visual comparisons between the PPD and measured data. \Cref{fig:ppd_dv} shows an example for DES~Y3 lensing ($\xi_+$ and $\xi_-$ shown together). In that Figure, the red points indicate real data and the orange band shows the PPD mean $\pm$ one standard deviation, computed using a Gaussian mixture PPD conditioned on that same cosmic shear data. Samples drawn from the PPD (not shown) reproduce the expected mean and variance, confirming consistency of the construction.

\begin{table}
    \centering
    \begin{tabular}{l p{1.4cm} p{1.4cm}}
        Test & {Quoted PPD $\tilde{p}$-value} & $\Delta_{\rm PPD}$ \newline (this work) \\
        \hline
        {\it redMaGiC (5 bins)}    & & \\
        $3\times2$ goodness-of-fit & 0.005 & 0.002 \\
        $2\times2$ goodness-of-fit & 0.038 & 0.004 \\
        $1\times2$ goodness-of-fit & 0.255 & 0.32 \\
        Shear \vs GGL+clustering   & 0.019 & 0.029 \\
        GGL \vs shear+clustering   & 0.016 & $<0.0001$ \\
        Clustering \vs shear+GGL   & 0.008 & $<0.0001$ \\[1ex]
        
        {\it MagLim (6 bins)}      &       & \\
        $3\times2$ goodness-of-fit & N/A   & $<0.0001$ \\
        $2\times2$ goodness-of-fit & $<0.001$ & $<0.0001$ \\
        $1\times2$ goodness-of-fit & 0.268 & 0.31 \\
        Shear \vs GGL+clustering   & N/A   & 0.36 \\
        GGL \vs shear+clustering   & N/A   & $<0.0001$ \\
        Clustering \vs shear+GGL   & N/A   & $<0.0001$ \\[1ex]
        
        {\it MagLim (4 bins)}      &       & \\
        $3\times2$ goodness-of-fit & 0.023 & 0.020 \\
        $2\times2$ goodness-of-fit & 0.019 & 0.021 \\
        $1\times2$ goodness-of-fit & 0.218 & 0.31 \\
        Shear \vs GGL+clustering   & 0.297 & 0.37 \\
        GGL \vs shear+clustering   & 0.004 & 0.002 \\
        Clustering \vs shear+GGL   & 0.526 & 0.32 \\
    \end{tabular}
    \caption{Comparison of PPD tests on DES~Y3 data. The first column indicates the specific test, while the second and third columns show results using the calibrated DES~Y3 metric and the new metric proposed in this work, $\Delta_{\rm PPD}$, respectively. Results are given for the two DES~Y3 lens samples, redMaGiC and MagLim (using either all six bins or with the two farthest bins removed). We evaluate goodness-of-fit for the $1\times2$ case (shear), $2\times2$ case (clustering plus galaxy–galaxy lensing, GGL), and $3\times2$ case (all three), as well as internal consistency tests where one two-point function is conditioned on the other two. When no PPD samples have a probability lower than the data, an upper bound ($<0.0001$) is reported.}
    \label{tab:delta_ppd_y3}
\end{table}

\Cref{tab:delta_ppd_y3} compares our new $\Delta_{\rm PPD}$ metric to the published DES~Y3 calibrated PPD $\tilde{p}$-values for three variations of DES Y3 joint analyses of lensing and clustering data. These analyses utilize different lens samples: redMaGiC (2.6 million galaxies in five tomographic bins) and MagLim (11 million galaxies in six bins, with the two farthest bins excluded in the final analysis).
Despite anticipated differences in numerical values, (since we do not calibrate $\Delta_{\rm PPD}$) the two metrics show remarkably consistent trends across all tests, demonstrating that our approach recovers known tensions while avoiding the numerical fragility of the Y3-method.

We consider in particular how these metrics respond to a property of the DES~Y3 {redMaGiC} galaxy sample captured by a systematics parameter,  $X_{\rm lens}$, applied as a multiplicative factor to the GGL signal. Effectively, it causes the GGL measurements to be modeled with different galaxy bias parameters than the galaxy clustering signal.  While this kind of scaling  is expected at small (halo-sized) scales due to the stochastic decorrelation between galaxy and matter distributions,  finding $X_{\rm lens}\neq 1$ at large scales is likely a sign of  systematic measurement errors. While the {redMaGiC} sample passed the DES Y3 unblinding requirements (PPD tests gave $\tilde{p}>0.01$ for shear and $2\times2$pt goodness of fit, as well as the disjoint test of shear vs $2\times2$pt), in the course of unblinding $2\times2$pt measurements were found to favor lower $S_8$ values than shear alone or $3\times2$pt, and additional post-unblinding PPD tests revealed low $\tilde{p}$-values.  Investigation revealed that the property driving this behavior was described by a stochasticity parameter with marginalized constraints of $X_{\rm lens}=0.88\pm0.02$, indicating a $\sim10\%$ suppression of the galaxy--galaxy lensing amplitude relative to the fiducial model, measured at the $5\sigma$ level. This suppression was traced to a color-dependent photometric issue in the {redMaGiC} galaxy sample, which was subsequently addressed by modifying the selection criteria~\citep*{y3-2x2ptbiasmodelling}. 
The fact that our new metric  $\Delta_{\rm PPD}$ and the calibrated Y3 $\tilde{p}$-value metric report a similarly low 3$\times$2pt goodness-of-fit, and a comparably good $1\times$2pt fit suggests it is responding to the same $X_{\rm lens}$-like decorrelation.  We also see that  $\Delta_{\rm PPD}$ gives an order-of-magnitude lower number for $2\times2$pt goodness-of-fit, and substantially lower values for the disjoint tests singling out either GGL or clustering. This is not necessarily a concern because these two metrics $\tilde{p}$ and $\Delta_{\rm PPD}$ are not expected to be identical. Rather, here we are concerned with whether low $\Delta_{\rm PPD}$ values would raise alerts to the same inconsistencies where there were low values for the Y3 $\tilde{p}$ metric.

For the MagLim sample, the DES~Y3 analysis initially used all six bins but ultimately adopted a more conservative approach by excluding the two highest-redshift bins to mitigate potential systematics and improve robustness.  The MagLim analysis also found evidence for a decorrelation similar to that seen for {redMaGiC}, but the use of fewer redshift bins and the optimized sample selection helped to reduce the impact of systematics, leading to more stable cosmological constraints~\citep*{y3-2x2ptaltlensresults}. As before, we see that where the Y3 $\tilde{p}$ reported low consistency, so does our new $\Delta_{\rm PPD}$ metric. Note that cases marked with `N/A' are PPD tests that were not reported in DES Y3 publications.

\section{Application to DES~Y6 unblinding}
\label{sec:y6}

The lessons from DES~Y3 directly inform the DES~Y6 unblinding protocol, where, among other tests, PPD tests are adopted as a fast and interpretable tool for assessing internal consistency across data subsets, as detailed in Section~IV and Appendix~A of~\citet*{y6-3x2pt}. In the DES~Y6 analysis, we use PPD to evaluate goodness-of-fit for the $3\times2$pt, $2\times2$pt, and $1\times2$pt combinations, as well as internal consistency tests in which one two-point function is conditioned on the other two. In the absence of formal calibration, we adopt a pragmatic interpretation of the PPD metric: $\Delta_{\rm PPD}$ values above $0.01$ are considered consistent, values below that trigger further investigation  and are assessed in the context of other diagnostics such as nuisance parameter posteriors and standard $\chi^2$ goodness-of-fit.  

Here we present a brief overview of the DES Y6 analysis in order to contextualize the simulated PPD tests presented below. Measurements are made from the DES Y6 Gold catalog~\citep*{y6-gold} built from the final DES Data Release 2~\citep{des-dr2}. 
We consider an updated MagLim lens sample split into six tomographic bins~\citep*{y6-maglim} using a combination of weights and masking to protect against observational systematics~\citep*{y6-mask}, and a source sample split into four bins~\citep*{y6-metadetect}, with point-spread function modeling done using the {\sc PiFF} software~\citep*{y6-piff}.  Cosmic shear, GGL, and the $2\times2$pt combination of GGL and galaxy clustering  are described in \citet*{y6-1x2pt}, \citet*{y6-gglens}, and \citet*{y6-2x2pt}, respectively. 

The DES~Y6 modeling framework builds upon Y3 with several important refinements which provide greater physical realism~\citep*{y3-generalmethods,y6-cardinal,y6-methods}. Nonlinear matter clustering is described with \textsc{HMCode2020}~\citep{2021MNRAS.502.1401M}, validated against the Euclid Emulator v2~\citep{1809.04695,2010.11288}, and baryonic effects are modeled via the parameter $\log T_{\rm AGN}$ fixed to a fiducial value of 7.7. Galaxy bias is treated using both a linear prescription and a nonlinear description based on Eulerian perturbation theory, implemented with \textsc{Fast-PT}~\citep{1603.04826,1609.05978} and validated against the \textsc{Aemulus-HEFT} simulations~\citep{2019ApJ...874...95Z}. In the nonlinear bias model, each of the six lens galaxy bins $i$ has two galaxy bias parameters, $b^{i}_1$ and $b^i_2$, corresponding to linear and quadratic bias terms in the expansion. In the linear bias model there is one galaxy bias parameter $b^i$ for each lens redshift bin. 
Additionally, the contribution of magnification to observed galaxy clustering is accounted for with one coefficient per lens bin~\citep*{y6-magnification}, with priors  based on synthetic source injections~\citep*{y6-balrog}. Intrinsic alignments are modeled using the TATT parameterization~\citep{1708.09247} in a simplified four-parameter framework.  Uncertainties over redshift distributions are marginalized over using mode-projection~\citep*{y6-nzmodes}, with priors set based on self-organizing map methods and cross-correlations with spectroscopic clustering measurements~\citep*{y6-sourcepz,y6-lenspz,y6-wz}. 
One multiplicative shear bias per source bin accounts for shear calibration uncertainties~\citep*{y6-metadetect}, with priors set by image simulation studies~\citep*{y6-imagesims}.   Scale cuts are validated iteratively to mitigate contamination from baryons, nonlinear clustering, and galaxy bias modeling approximations~\citep*{y6-methods}.

\subsection{Simulated Y6 analysis}

To validate our PPD methodology for DES~Y6, we apply it to simulated analyses. Unlike most simulated analyses used to validate the DES~Y6 modeling pipeline, tests of the PPD metric cannot rely on noiseless synthetic data, since by definition the PPD quantifies the statistical consistency between noisy realizations of the data and the model. We therefore generate ten Gaussian noise realizations drawn from the analytic data covariance of~\citep*{y6-methods} and add them to two model predictions evaluated at the fiducial cosmology: one unperturbed, and one with an injected internal inconsistency corresponding to a GGL-vs-clustering decorrelation of $X_{\rm lens}=0.9$. Based on the marginalized constraints from a simulated analysis with free $X_{\rm lens}$, we find that  value produces an effective $5\sigma$ deviation in parameter space relative to $\Lambda$CDM ($X_{\rm lens}=1$).

To make the analysis of ten noisy realizations computationally feasible, we employ a fast, approximate version of the DES~Y6 $3\times2$pt pipeline. This version replaces the full nonlinear matter power spectrum with the Eisenstein \& Hu linear approximation~\citep{1999ApJ...511....5E}. Also, because this fast pipeline was developed prior to the finalization of the fiducial Y6 nuisance parameter priors and scale cuts, it uses the simpler shift–stretch parameterization of photometric redshift uncertainties (as in Y3) and slightly different angular scale cuts than the final configuration. We consider versions of this analysis with both linear and nonlinear galaxy bias, employing different scale cuts to account for the fact that  nonlinear bias modeling is accurate to smaller angular scales. The simulated analyses use normalized redshift distributions based on DES~Y3 samples, but the covariance matrix is updated to reflect the larger area and reduced shape noise of Y6. Magnification bias is fixed for each lens bin, rather than varied with a Gaussian prior as in the final Y6 setup, and the covariance does not yet include the large-scale structure weights correction applied in the final analysis. Overall, our simulated DES Y6 $3\times2$pt data vector with linear (nonlinear) bias consists of a total of 653 (669) data points, which are modeled in flat $\Lambda$CDM with 6 cosmological parameters including neutrino mass, 6 (12) galaxy bias parameters, and 24 other nuisance parameters describing intrinsic alignments, shear calibration, and photo-$z$ uncertainties. Many of the nuisance parameters have Gaussian priors informed by calibration and sample characterization studies.  Because this  fast pipeline approximately preserves the parameter space dimensionality and modeling degrees of freedom of the full Y6 pipeline, we expect  these PPD to be broadly representative of performance for the real data analysis. However, as we will discuss in \cref{sec:summary-discussion},  the fact that differences between these pipelines may influence the detailed behavior of the PPD metric was a consideration when designing the DES Y6 unblinding protocol. 

We use this pipeline to rehearse the PPD-relevant component of the DES~Y6 unblinding protocol across three model variants: (i) a nonlinear bias model, (ii) a linear bias model, and (iii) a linear bias model with fixed nuisance parameters. For each variant, we compute PPD metrics for both the fiducial data vector and the one with injected $X_{\rm lens}$ tension,\footnote{We also tested a vector contaminated with baryonic feedback, but found that because scale cuts largely protect the analysis against it, it yielded  no significant effect.} sampling posteriors across all ten noise realizations to evaluate $\Delta_{\rm PPD}$.

\subsection{Results}

\begin{figure*}
    \centering
    \includegraphics[scale=0.85]{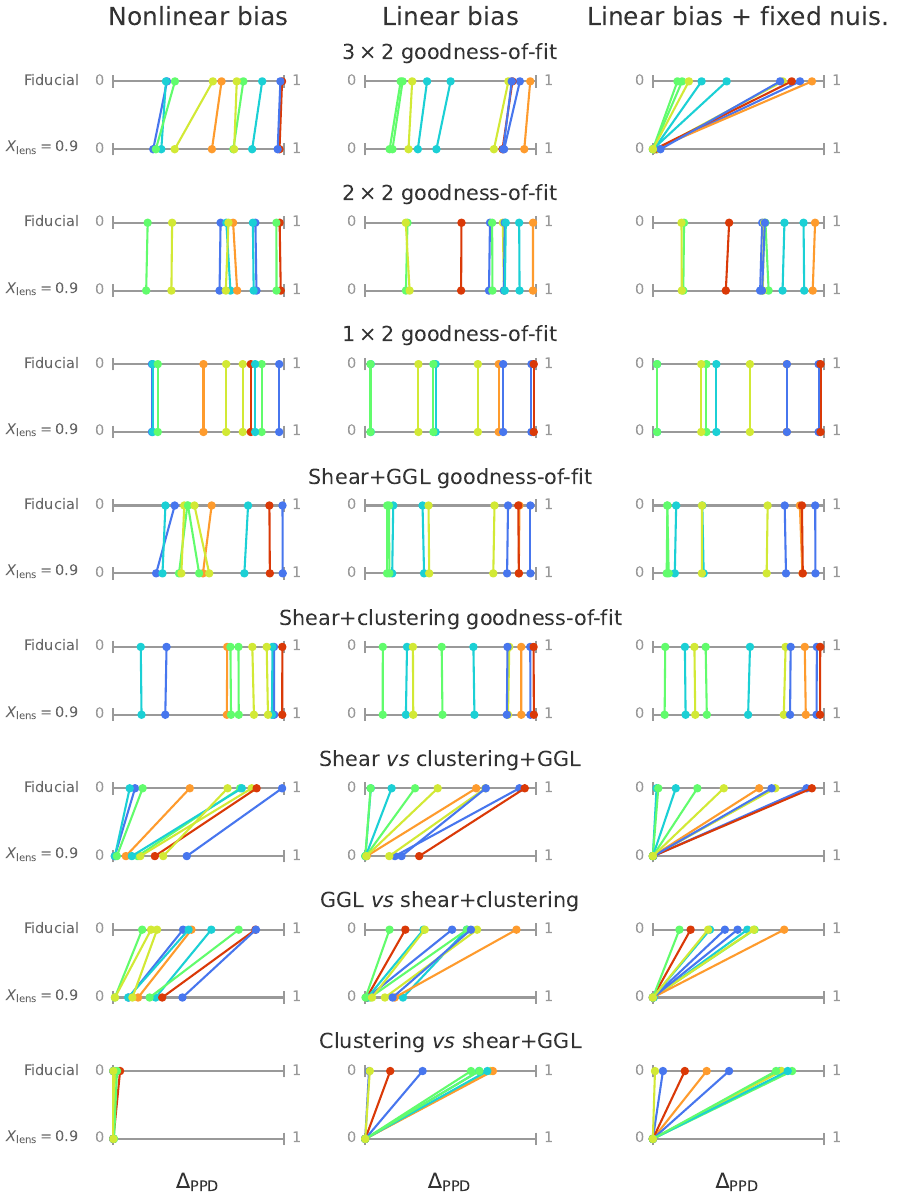}
    \caption{PPD tests for DES~Y6 simulated data under different model variations. Each column corresponds to a different modeling choice: nonlinear galaxy bias (left), linear bias (middle), and linear bias with fixed nuisance parameters (right). Each row corresponds to a different test, with goodness-of-fit tests shown in the top five rows and consistency tests between disjoint data subsets in the bottom three rows. Individual panels display $\Delta_{\rm PPD}$ computed from multiple noise realizations (one color per realization), comparing the fiducial data vector (top of each panel) to a modified data vector with $X_{\rm lens}=0.9$ (bottom).    We see that fixing nuisance parameters in the linear bias model (right column) yields the strongest shifts in $\Delta_{\rm PPD}$ towards inconsistency in the $3\times2$ goodness-of-fit test (top row) and disjoint consistency tests (bottom three rows). This sensitivity decreases as the model becomes more flexible, with the nonlinear bias model (left column) yielding much smaller shifts, which we interpret as $X_{\rm lens}$ being partially absorbed by the additional bias and nuisance parameters.
    }
    \label{fig:y6-ppd}
\end{figure*}

The results of our simulated DES~Y6 analyses are  summarized in \cref{fig:y6-ppd}. In that figure, the three columns correspond to different modeling choices, with the most complex nonlinear galaxy bias model on the left, linear bias in the middle, and a version of the linear bias model with fixed nuisance parameters on the right. Each row shows a different PPD test configuration, with sub-panels showing how the inferred $\Delta_{\rm PPD}$ changes when evaluated for simulated data with and without an injected $X_{\rm lens}$ inconsistency. Each set of colored markers corresponds to one of the ten noise realizations studied.

We quantify  the shift in $\Delta_{\rm PPD}$, defined as the difference between the mean $\Delta_{\rm PPD}$ values for the fiducial and $X_{\rm lens}=0.9$ data vectors, normalized by their standard deviations over the ten noise realizations added in quadrature:
\begin{equation}\label{eq:Deltashift}
    {\rm shift} = 
    \frac{
        \bar{\Delta}_{\rm PPD}^{(X_{\rm lens}=0.9)} - \bar{\Delta}_{\rm PPD}^{\mathrm{(fid)}}
    }{
        \sqrt{
            (\sigma_{X_{\rm lens}})^2 + (\sigma_{\rm fid})^2
        }}.
\end{equation}
The sign convention is such that negative values indicate reduced PPD consistency for the contaminated ($X_{\rm lens}=0.9$) data vector relative to the fiducial case. The resulting shifts for all goodness-of-fit and consistency tests, and for each modeling choice considered, are reported  in \cref{tab:ppdshifts}. 

\begin{table}
    \centering
    \begin{tabular}{p{3.2cm} p{1.0cm} p{1.0cm} p{1.4cm}}
        Test & Nonlinear bias & Linear bias & Linear bias \newline + fixed nuis. \\
        \hline
        $3\times2$ goodness-of-fit & -0.2  & -0.1  & -1.7 \\
        $2\times2$ goodness-of-fit & 0.0  & 0.0  & -0.0 \\
        $1\times2$ goodness-of-fit & 0.0  & 0.0  & 0.0 \\
        Shear+GGL goodness-of-fit & 0.0  & 0.0  & 0.0 \\
        Shear+clustering GoF & 0.0  & 0.0  & 0.0 \\
        Shear \textit{vs} clustering+GGL & -1.2  & -1.2  & -1.4 \\
        GGL \textit{vs} shear+clustering & -1.3  & -1.8  & -2.5 \\
        Clustering \textit{vs} shear+GGL & -0.5  & -1.6  & -1.6 \\
    \end{tabular}
    \caption{Quantitative summary of the $\Delta_{\rm PPD}$ shifts shown in \cref{fig:y6-ppd}. The table reports the shifts in $\Delta_{\rm PPD}$, as defined in \cref{eq:Deltashift}, for the DES~Y6 simulated data tests under the three model variations considered: nonlinear bias, linear bias, and linear bias with fixed nuisance parameters. Each entry corresponds to the change in $\Delta_{\rm PPD}$ induced by setting $X_{\rm lens}=0.9$, for goodness-of-fit tests (top five rows) and consistency tests between disjoint data subsets (bottom three rows). As seen in \cref{fig:y6-ppd}, the largest shifts appear in cross-probe consistency tests and are enhanced when nuisance parameters are fixed.}
    \label{tab:ppdshifts}
\end{table}

To aid in the interpretation of these results, it is helpful to consider how the observed two-point functions respond to different parameter combinations. If $b$ labels generic galaxy bias parameters, shear signals scale like $\sigma_8^2$ and are independent of galaxy bias, GGL signals scale like the product $X_{\rm lens}\,b\sigma_8^2$, and galaxy clustering scales like $b^2\sigma_8^2$. Given these dependencies, we neither expect nor observe goodness-of-fit PPD tests for any pair of these correlation functions to be sensitive to an injected $X_{\rm lens}$ inconsistency. 
In contrast, we do expect $X_{\rm lens}$ to impact goodness-of-fit tests for the full $3\times 2$pt measurement and to impact disjoint tests of one probe conditioned on the other two.  Our findings are aligned with this expectation:  the injected  inconsistency tends to lower $\Delta_{\rm PPD}$ for data combinations that are able to break degeneracies between $\sigma_8$, $X_{\rm lens}$, and galaxy bias. While most $\Delta_{\rm PPD}$ values decrease with the injected $X_{\rm lens}$ inconsistency, a few instances show an increase; this behavior can arise from statistical fluctuations in the data noise, finite posterior sampling, or the way nuisance parameters interact with their prior boundaries, although we do not attribute these cases to a single specific effect.

However, while $\Delta_{\rm PPD}$ generally performs as expected, these tests also reveal two counterintuitive behaviors that inform the metric's practical application. First, we note that the $3\times 2$pt  goodness-of-fit tests are less sensitive to contamination than one might expect given the $5\sigma$ parameter shift of the injected $X_{\rm lens}=0.9$. Second, we find that for the  nonlinear galaxy bias model, the disjoint PPD test of galaxy clustering conditioned on GGL plus shear reports inconsistency even for internally consistent data (see the plot in the bottom-left of \cref{fig:y6-ppd}).  We trace these behaviors to different manifestations of the model's flexibility, as described below.

\subsubsection{Nuisance parameters can absorb data contamination.}\label{sec:nuis-absorb}

The $3\times2$pt goodness-of-fit tests show surprisingly weak responses to the injected tension, with shifts of approximately $-0.2$ for the nonlinear bias model and $-0.1$ for the linear model. We cross check this PPD test with  the best-fit $\chi^2$, roughly estimated from posterior samples,\footnote{Chains used to infer posteriors lack the sample density needed for a precise estimate of the best-fit. This motivates additional optimizer runs as described in Appendix~H of~\citet*{y6-methods} when more precise evaluations are needed for model comparison, but chain-based estimates are sufficient for our purposes here.} finding the presence of the $X_{\rm lens}$ contamination increases the minimum $\chi^2$ by only $\Delta\chi^2 \sim 5.5\pm2.8$ (mean and standard deviation over the ten realizations) for the contaminated versus fiducial  $3\times2$pt data vector. Given the size of the data vector (corresponding to hundreds of degrees of freedom), this is a modest response to a nominally $5\sigma$ injected inconsistency. Inspection of the nuisance parameter posteriors reveals why: several lens photo-$z$ width parameters shift $1-2\sigma$ below their Gaussian priors, particularly at low redshift, allowing the model to partially compensate for the $X_{\rm lens}$ mismatch by boosting the clustering signal relative to GGL. The model thus absorbs the injected inconsistency through its nuisance degrees of freedom rather than expressing it in $\Delta_{\rm PPD}$. This is confirmed by the fixed-nuisance tests shown in the third column of \cref{fig:y6-ppd} and \cref{tab:ppdshifts}, in which the $3\times2$pt goodness-of-fit shifts by $-1.7$. 

The fact that inconsistencies are being absorbed by nuisance parameters does not mean that we are insensitive to them, just that they cannot be detected with a PPD-based metric.  Nuisance parameter priors are informed by other aspects of measurements (e.g. photo-$z$ calibration), so if a nuisance parameter posterior is offset relative to its prior, this indicates the presence of some effect that is not well described by the model or assumptions about the data. These results underscore the importance of examining nuisance parameter posteriors alongside PPD metrics during unblinding.

Beyond the muted sensitivity due to nuisance parameters, this behavior also highlights an intrinsic property of goodness-of-fit tests in high-dimensional data spaces. A $5\sigma$ injected inconsistency means that a model comparison adding $X_{\rm lens}$ as a free parameter would show a $\sim 5\sigma$ preference over the fixed-$X_{\rm lens}$ model, corresponding to $\Delta\chi^2\sim 25$ at the best fit. In contrast, goodness-of-fit tests assess $\chi^2$ relative to the number of degrees of freedom, and the probability-to-exceed becomes increasingly insensitive to moderate $\chi^2$ changes as the number of data points grows~\citep{2019PhRvD..99d3506R}. For our simulated DES~Y6 analyses with $\sim 600$ degrees of freedom, the standard deviation of $\chi^2$ is $\sqrt{N_{\rm d.o.f.}}\approx 25$, so $\Delta\chi^2\sim 25$ corresponds to only a $\sim 1\sigma$ shift in the goodness-of-fit. In other words, goodness-of-fit tests are less sensitive to a given feature than model comparison tests, which measure how significantly that feature is captured by a single additional parameter.

We expect this to be a general property of goodness-of-fit tests, including those based on PPDs. A benefit of this property is that even if PPD-based tests are conducted for a $\Lambda$CDM cosmological model, they are unlikely to report inconsistency simply due to the data preferring an alternative cosmological model, such as dynamical dark energy. In other words:  using PPD tests as part of unblinding checks can alert us to significant issues with the analysis pipeline, it presents very little risk of artificially guiding the analysis towards $\Lambda$CDM results.

\subsubsection{Projection effects can skew PPDs.}\label{sec:projection}

\begin{figure*}
    \centering
    \includegraphics[scale=0.65]{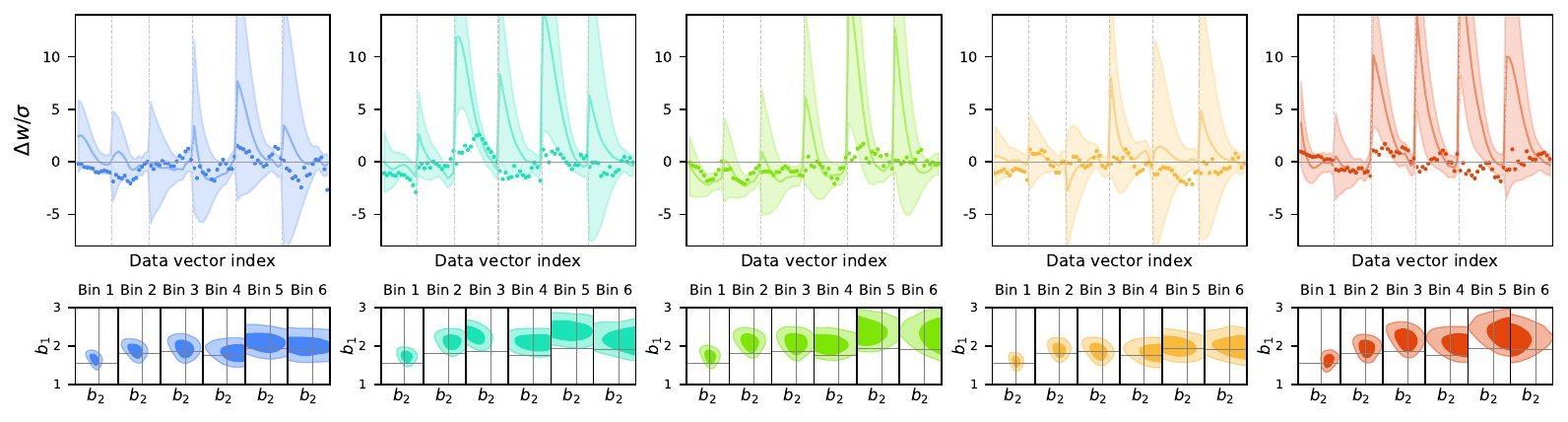}
    \caption{Illustration of projection effects in simulated DES~Y6 internal consistency tests with the nonlinear bias model. Five randomly selected realizations (out of the 10  used in \cref{fig:y6-ppd}) of the full $3\times2$pt data vector are shown, each in a different color. Top: clustering measurements $w(\theta)$ (colored points) compared to the PPD band predicted from shear and galaxy-galaxy lensing, both shifted by the fiducial input model and normalized by their standard deviations; vertical grey lines mark the boundaries of the tomographic lens bins. Bottom: marginalized posteriors of the linear and nonlinear bias parameters, respectively, inferred from shear and galaxy-galaxy lensing in each lens bin, with grey lines indicating the input values. The vertical extent corresponds to the uniform prior ranges, [1,3] for $b_1$ and [-3,+3] for $b_2$.}
    \label{fig:y6_wtheta_sg}
\end{figure*}

A distinct  phenomenon appears in internal consistency tests that condition clustering on shear and GGL, particularly for the nonlinear bias model: $\Delta_{\rm PPD}$ takes anomalously low values even for fiducial, internally consistent data. We posit that the origin of this behavior lies in projection effects related to parameter degeneracies. 

Projection effects (sometimes called prior-volume effects) occur in Bayesian analyses when unconstrained directions in higher-dimensional parameter space cause marginalized (projected) posteriors to be offset from best-fit parameter values.  As large-scale structure analyses have grown in complexity, projection effects have become increasingly prevalent in recent years, motivating  careful characterization using simulated analyses \citep*{y6-methods,2411.12021}, complementary frequentist  methods~\citep[e.g.~][]{2408.07700,2409.09101,2506.12004,2108.11518,2403.08101}, as well as approaches using simulation-based priors~\citep{2511.20757}.

How is this related to the observed PPD behavior? Since the PPD is the posterior convolved with the likelihood, it inherits any projection effects impacting the marginalized posteriors, thus breaking the intuition that unconstrained parameters should produce broad PPDs and favor consistency, as discussed in \cref{sec:metric}. In most cases this is not a concern, because projection effects tend to have less impact on parameters more directly related to observables---which is what motivates, for example, sampling products of galaxy bias and $\sigma_8$ rather than galaxy bias alone~\citet{2411.12022}. This intuition does not hold, however, for PPD tests between disjoint observables sensitive to different degenerate parameter combinations.

Though combining GGL with shear allows us to break degeneracies between galaxy bias and $\sigma_8$, for the nonlinear bias model this data combination is only able to constrain degenerate combinations of galaxy bias parameters, leaving individual bias parameters poorly determined, particularly at high redshift. Combined effects from noise, degeneracies with nuisance  parameters, and prior boundaries can then produce offsets in marginalized posteriors. Since clustering is sensitive to the square of bias parameters, these offsets are enhanced when translating to predictions for the galaxy-galaxy angular two-point correlation function $w(\theta)$, shifting the PPD away from the true data realization and producing low $\Delta_{\rm PPD}$ even for consistent data (which we observe in the left bottom plot of \cref{fig:y6-ppd}).
The disjoint PPD test of GGL conditioned on shear plus clustering is less sensitive to this behavior because of galaxy clustering's stronger sensitivity to the bias parameters. We would expect that a linear bias model would be less sensitive to this behavior, though the $b$ versus $b^2$ sensitivity of GGL and clustering respectively could still potentially skew the disjoint PPD if the GGL-plus-shear bias posteriors are impacted by degeneracies with other nuisance parameters. 

To support this intuition,
\cref{fig:y6_wtheta_sg} shows the galaxy-clustering  PPDs for a subset of data vector realizations (upper panels) alongside marginal distributions of the linear ($b_1$) and nonlinear ($b_2$) bias parameters for each lens bin (lower panels). We find that for certain realizations and bins, the posteriors appear somewhat discrepant with the input bias parameter values, with $b_1$ found consistently high (though still within $2\sigma$), whereas $b_2$ is skewed towards the prior limits. This propagates into offsets of the predicted clustering signal in the upper panels, especially at small scales where $b_2$ contributes significantly (towards the left between vertical lines separating tomographic bins). This particularly affects high redshift bins (right end of each panel), where lower signal-to-noise translates to weaker constraints on the galaxy bias parameters. Switching to a simpler  linear bias model tends to mitigate this effect: posteriors recover the input $b_1$ values and the PPD bands better capture the behavior of the input model at all scales.   This figure also illustrates the fact that projection effects can only shift the PPD away from the true data realization when parameter uncertainty dominates the data covariance (in the sense of \cref{eq:ppd_cov}), because it is only then that offsets in marginalized posteriors translate into visible shifts in the PPD rather than being washed out by data noise. In \cref{fig:y6_wtheta_sg}, the PPD is substantially wider than the likelihood at the input model, especially at small scales and high redshifts, confirming that parameter uncertainty dominates the PPD width.

These observations suggest a simple mitigation strategy. When a flexible model is used, PPD tests should be validated with a few realizations of simulated analyses to check for projection effects that might drive $\Delta_{\rm PPD}$ towards anomalously low values---falsely indicating inconsistency---even for perfectly compatible data. Unblinding  criteria can then be adjusted to discount PPD test configurations subject to these  effects. This is precisely how PPD tests were deployed in the DES Y6 analysis. Of course, one could perform a large ensemble of simulated analyses to calibrate the size of projection effects, thus translating $\Delta_{\rm PPD}$ to a formal $p$-value. However, as a practical consideration, running simulated analyses for just a few noise realizations and adjusting validation criteria accordingly is much more feasible within the workflow of a cosmological  analysis.

\section{Summary and discussion}
\label{sec:summary-discussion}

In this paper we introduced a fast and interpretable posterior-predictive metric, $\Delta_{\rm PPD}$, for conducting internal consistency tests of correlated cosmological probes. Defined as the fraction of the posterior predictive distribution (PPD)---approximated as a Gaussian mixture built from posterior samples---at lower probability density than the observed data, it compresses the high-dimensional PPD into an interpretable scalar that can be computed in minutes from existing posterior chains.
We emphasize that $\Delta_{\rm PPD}$ is a diagnostic indicator rather than a calibrated $p$-value.  This reflects the hybrid nature of PPD tests, which pose a frequentist consistency question using Bayesian 
machinery: the data-sampling scheme defining the null hypothesis is not unique, and  as demonstrated in \cref{app:pvalue},  $\Delta_{\rm PPD}$ is not uniformly distributed under the schemes we 
consider. Obtaining a formal $p$-value would therefore require choosing 
a scheme and calibrating against a large ensemble of simulated analyses dedicated solely to characterizing its sampling distribution. Given the significant associated computational cost, for practical applications we instead recommend running a few simulated realizations sufficient to validate PPD behavior, which are in any case useful for broader pipeline and model validation. 

Applied to DES~Y3 data, the uncalibrated metric reproduces the qualitative trends of the calibrated PPD $\tilde{p}$-values reported in DES~Y3 analyses (\cref{tab:delta_ppd_y3}), identifying the same tensions between probes while avoiding the numerical fragility and computational overhead of DES-Y3-like importance-sampling calibration. We also demonstrate that this PPD method yields intuitive data-space visual checks (\cref{fig:ppd_dv}).

For the simulated DES~Y6 analyses, we carried out PPD consistency tests on ten noisy realizations of synthetic $3\times2$pt data, using both internally consistent realizations and realizations with an injected decorrelation between GGL and clustering ($X_{\rm lens}=0.9$). We considered three modeling variants: a nonlinear galaxy bias model, a linear bias model, and a linear bias model with fixed nuisance parameters.
These tests, reported in \cref{fig:y6-ppd}, allowed us to identify regimes where $\Delta_{\rm PPD}$ performs as expected, as well as two apparent failure modes that underscore the need to combine PPD tests with other unblinding diagnostics. In some cases, $\Delta_{\rm PPD}$ remains high despite an injected inconsistency, highlighting that model flexibility can absorb genuine tensions---as we see  when lens photo-$z$ width parameter estimates drift away from their Gaussian priors to compensate for the injected decorrelation. In such cases, pre-unblinding checks of nuisance parameter posteriors serve as important complements to PPD tests. Conversely, low $\Delta_{\rm PPD}$ values for consistent simulated data flag potential projection effects due to unconstrained directions in parameter space---as we saw for disjoint tests conditioned on GGL-plus-shear, where poorly constrained bias parameters distort the predicted clustering PPD. This highlights the fact that applications to simulated analyses should guide how PPD tests are used as pre-unblinding checks in a real analysis. In some cases, the study of how projection effects impact the PPD metric may aid in identifying aspects of the modeling that warrant refinement, such as reducing model complexity or tightening prior choices. 

These findings directly informed the DES~Y6 unblinding protocol, in which PPD tests of real DES~Y6 data were used to validate the analysis before cosmological results were revealed and are reported in~\citet*{y6-1x2pt,y6-2x2pt,y6-3x2pt}. The Y6 protocol used $\Delta_{\rm PPD}>0.01$ as a threshold for passing consistency tests, with lower values triggering further scrutiny of the analysis pipeline. Goodness-of-fit PPD tests were applied to shear, $2\times 2$pt (GGL+clustering) and $3\times 2$pt for several model variations, as were disjoint tests for shear conditioned on $2\times 2$pt and GGL conditioned on shear-plus-clustering. Based on the projection effects seen in~\cref{sec:y6} above, the disjoint test of clustering conditioned on GGL-plus-shear was not used\footnote{While in this work we only saw significant projection effects for the nonlinear galaxy bias model, in the real Y6 analysis this disjoint test was neglected for both linear and nonlinear  bias. The fiducial DES~Y6 analysis uses a mode-projection~\citep*{y6-nzmodes} parameterization of photometric redshift uncertainties as opposed to the shift-stretch model (from Y3) applied here, and mode projection introduces new nuisance parameters that could potentially contribute to projection effects.  In the absence of explicitly simulated PPD checks, we opt to neglect these tests to avoid false positives in checks for inconsistency}
and  low $\Delta_{\rm PPD}$ found for nonlinear bias fits to GGL conditioned on shear plus clustering were not treated as a reason to halt the analysis. Notably, the goodness-of-fit PPD tests, supported by scrutiny of nuisance parameter posteriors, were instrumental in identifying a problem with the second lens redshift bin, which was ultimately removed from the fiducial analysis of~\citet*{y6-3x2pt}.

 Looking ahead, the framework introduced here is well suited for application to forthcoming Stage-IV surveys such as LSST, Euclid, and Roman. Their increased statistical power will make fast, robust internal consistency tests essential, while their more complex models will require a clear understanding of how these metrics couple to modeling choices. This represents a crucial task in the broader project of precision large-scale structure cosmology, but one that this work shows we can carry out effectively.

\section*{Author Contributions}

All authors contributed to this paper and/or carried out infrastructure work that made the analysis possible. C.~Doux and E.~Baxter developed the initial concept of the Gaussian mixture PPD approximation and the $\Delta_{\rm PPD}$ metric. C.~Doux created a prototype code implementation used for proof-of-concept tests and carried out the toy model demonstrations of PPD properties. J.~Muir led the implementation of the PPD method within the DES Y6 analysis framework,  the simulated analyses supporting its validation, and its application within the real DES Y6 analysis, with support from A.~Fert\'e and D. Sanchez Cid. M.~Raveri and N.~Weaverdyck contributed insight aiding the interpretation of the simulated DES Y6 test results.  C.~Doux and J.~Muir contributed to the writing of the manuscript. T.~Karwal and W.~d'Assignies served as the internal review committee and provided valuable feedback on the writing and presentation of this paper. M.~A.~Troxel, C.~Chang, M.~Crocce, and M.~R.~Becker contributed to the development of this work as coordinators of the DES Year 6 analysis. The remaining authors have made contributions to this paper that include, but are not limited to the construction of DE-Cam and other aspects of collecting data: data processing and calibration; developing broadly used methods, codes, and simulations; running the pipeline and validation tests; and promoting the science analysis.

\section*{Data Availability}

The DES Y6 data used in this analysis, including catalogues, data-vector measurements, likelihoods and chains, are publicly available at \href{https://des.ncsa.illinois.edu/releases/y6a2}{\texttt{https://des.ncsa.illinois.edu/releases/y6a2}}.

The code implementing the PPD method presented in this work is publicly available at \href{https://github.com/des-science/y6-ppd-public/}{\texttt{https://github.com/des-science/y6-ppd-public/}}.

\section*{Acknowledgements}

Funding for the DES Projects has been provided by the U.S. Department of Energy, the U.S. National Science Foundation, the Ministry of Science and Education of Spain, 
the Science and Technology Facilities Council of the United Kingdom, the Higher Education Funding Council for England, the National Center for Supercomputing 
Applications at the University of Illinois at Urbana-Champaign, the Kavli Institute of Cosmological Physics at the University of Chicago, 
the Center for Cosmology and Astro-Particle Physics at the Ohio State University,
the Mitchell Institute for Fundamental Physics and Astronomy at Texas A\&M University, Financiadora de Estudos e Projetos, 
Funda{\c c}{\~a}o Carlos Chagas Filho de Amparo {\`a} Pesquisa do Estado do Rio de Janeiro, Conselho Nacional de Desenvolvimento Cient{\'i}fico e Tecnol{\'o}gico and 
the Minist{\'e}rio da Ci{\^e}ncia, Tecnologia e Inova{\c c}{\~a}o, the Deutsche Forschungsgemeinschaft and the Collaborating Institutions in the Dark Energy Survey. 

The Collaborating Institutions are Argonne National Laboratory, the University of California at Santa Cruz, the University of Cambridge, Centro de Investigaciones Energ{\'e}ticas, 
Medioambientales y Tecnol{\'o}gicas-Madrid, the University of Chicago, University College London, the DES-Brazil Consortium, the University of Edinburgh, 
the Eidgen{\"o}ssische Technische Hochschule (ETH) Z{\"u}rich, 
Fermi National Accelerator Laboratory, the University of Illinois at Urbana-Champaign, the Institut de Ci{\`e}ncies de l'Espai (IEEC/CSIC), 
the Institut de F{\'i}sica d'Altes Energies, Lawrence Berkeley National Laboratory, the Ludwig-Maximilians Universit{\"a}t M{\"u}nchen and the associated Excellence Cluster Universe, 
the University of Michigan, NSF NOIRLab, the University of Nottingham, The Ohio State University, the University of Pennsylvania, the University of Portsmouth, 
SLAC National Accelerator Laboratory, Stanford University, the University of Sussex, Texas A\&M University, and the OzDES Membership Consortium.

Based in part on observations at NSF Cerro Tololo Inter-American Observatory at NSF NOIRLab (NOIRLab Prop. ID 2012B-0001; PI: J. Frieman), which is managed by the Association of Universities for Research in Astronomy (AURA) under a cooperative agreement with the National Science Foundation.

The DES data management system is supported by the National Science Foundation under Grant Numbers AST-1138766 and AST-1536171.
Data access is enabled by Jetstream2 and OSN at Indiana University through allocation PHY240006: Dark Energy Survey from the Advanced Cyberinfrastructure Coordination Ecosystem: Services and Support (ACCESS) program, which is supported by U.S. National Science Foundation grants 2138259, 2138286, 2138307, 2137603, and 2138296.
The DES participants from Spanish institutions are partially supported by MICINN under grants PID2021-123012, PID2021-128989 PID2022-141079, SEV-2016-0588, CEX2020-001058-M and CEX2020-001007-S, some of which include ERDF funds from the European Union. IFAE is partially funded by the CERCA program of the Generalitat de Catalunya.

We  acknowledge support from the Brazilian Instituto Nacional de Ci\^encia
e Tecnologia (INCT) do e-Universo (CNPq grant 465376/2014-2).

This document was prepared by the DES Collaboration using the resources of the Fermi National Accelerator Laboratory (Fermilab), a U.S. Department of Energy, Office of Science, Office of High Energy Physics HEP User Facility. Fermilab is managed by Fermi Forward Discovery Group, LLC, acting under Contract No. 89243024CSC000002.

Some calculations were done using the University of Cincinnati's Advanced Research Computing center as well as the Symmetry cluster at Perimeter Institute. Research at Perimeter Institute is supported in part by the Government of Canada through the Department of Innovation, Science and Economic Development and by the Province of Ontario through the Ministry of Colleges and Universities.

This work made use of the following software packages:
\textsc{NumPy}~\citep{Harris2020},
\textsc{Matplotlib}~\citep{Hunter2007},
\textsc{SciPy}~\citep{2020SciPy-NMeth},
\textsc{TensorFlow}~\citep{Abadi2016},
\textsc{TensorFlow Probability}~\citep{1711.10604},
\textsc{Nautilus}~\citep{2024AAS...24422902L}, and
\textsc{CosmoSIS}~\citep{cosmosis2020}.

\bibliographystyle{yahapj_twoauthor}

\bibliography{example,y3kp,y6kp,des}

\begin{appendix}

\section{Is $\Delta_{\rm PPD}$ a $p$-value?}
\label{app:pvalue}

\begin{figure*}
    \centering
    \begin{minipage}{0.49\textwidth}
        \centering
        \begin{tikzpicture}
            \node at (0,0) {\includegraphics[scale=0.65]{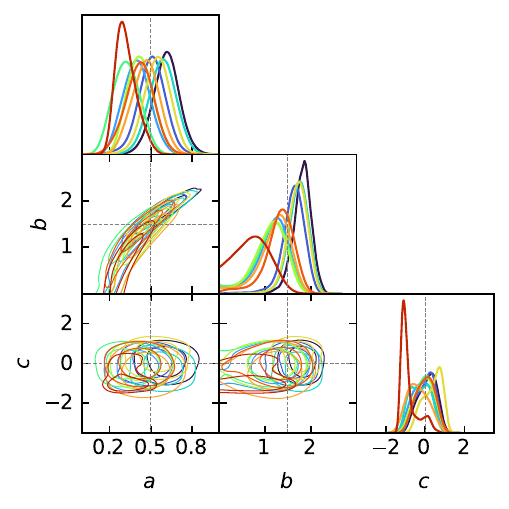}};
            \node at (3.2,1.15) {\includegraphics[scale=0.65]{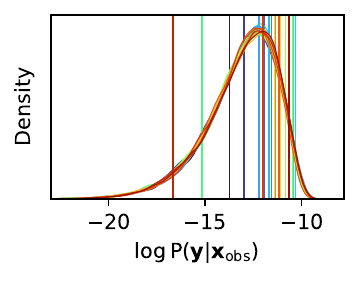}};        
        \end{tikzpicture}
    
        \small (a) $\theta=\theta_0$ test.
    \end{minipage}
    \begin{minipage}{0.49\textwidth}
        \centering
        \begin{tikzpicture}
            \node at (0,0) {\includegraphics[scale=0.65]{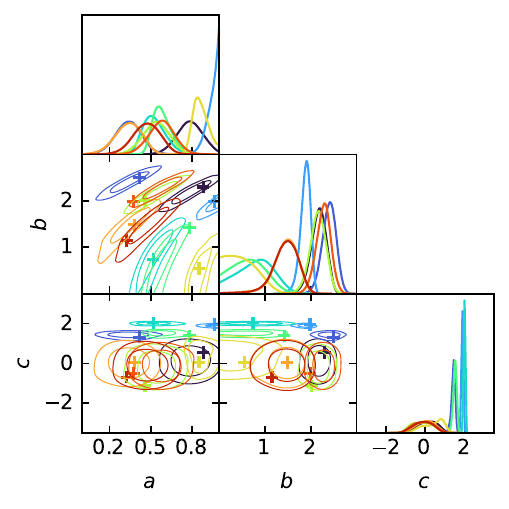}};
            \node at (3.2,1.15) {\includegraphics[scale=0.65]{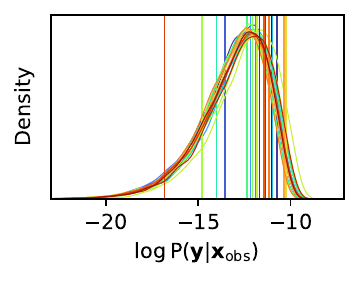}};        
        \end{tikzpicture}
    
        \small (b) $\theta\sim{\rm P}(\theta)$ test.
    \end{minipage}
    \caption{Toy model PPD tests under two frameworks. Left: frequentist approach, where all data realizations are generated from the same input parameters. Right: Bayesian approach, where each data realization is generated from parameters drawn from the prior. In both cases, the corner plot illustrates a subset of posterior distributions (parameters values indicated by grey lines on the left and  “+” symbols on the right). The upper-right panels show the histograms of log-density values for PPD samples compared to the data realizations.}
    \label{fig:toy-ppd2}
\end{figure*}

\begin{figure}
    \centering
    \includegraphics[scale=0.65]{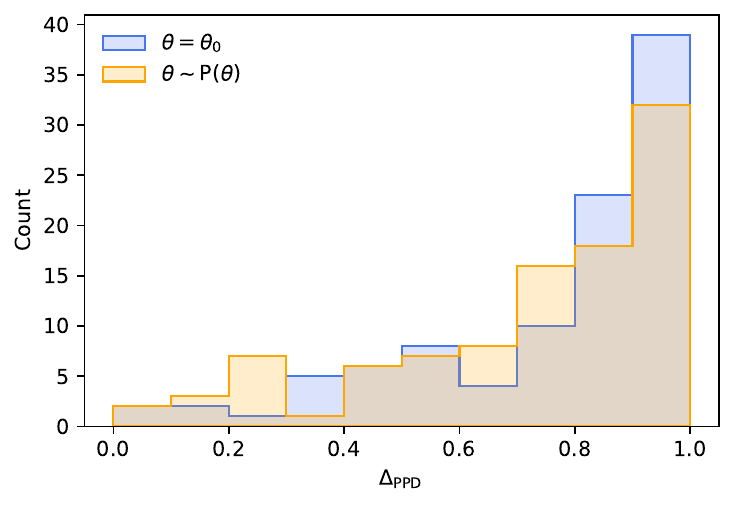}
    \caption{Histograms of $\Delta_{\rm PPD}$ values from 100 realizations, comparing the frequentist (blue) and Bayesian (orange) approaches. In both cases the distributions deviate from uniformity.}
    \label{fig:toy-ppd3}
\end{figure}

Assuming that the statistical description of $\vd$ and $\vdp$ is accurate, \ie that the likelihood correctly describes the random process that produced the data, then answering this question really amounts to deciding which distribution $\vdpobs$ is drawn from. 
This ambiguity is a direct consequence of the hybrid nature of the test: asking whether the observed data are typical of the model is a frequentist question, which requires specifying a data-sampling scheme that Bayesian inference alone does not uniquely define.

Under the hypothesis that the observed $\vdpobs$ is itself a realization drawn, \textit{a posteriori}, from the posterior predictive distribution defined for $\vdobs$, then $\Delta_{\rm PPD}$ is, by construction, uniformly distributed in the range $[0,1]$ over realizations of $\vdp$, \ie it is a $p$-value.

If, instead, we were to average $\Delta_{\rm PPD}$ over realizations of $\vdpobs$ drawn by first sampling the prior on $\bth$ and then the likelihood $\Prob{\vdobs,\vdpobs|\bth}$ simultaneously with $\vdobs$, ahead of inferring the posterior $\Prob{\bth | \vdobs}$, then $\Delta_{\rm PPD}$ would, in general, not be uniformly distributed. Both hypotheses seem reasonable in a Bayesian framework. 

In a frequentist context, where $\vdobs$ and $\vdpobs$ are assumed to be drawn simultaneously from the likelihood $\Prob{\vdobs,\vdpobs|\bth_0}$ at the true (but unknown) value of the model parameters $\bth_0$, then repeating the test over such data realizations would, in general, not yield a uniform distribution.

To verify this, we repeat PPD goodness-of-fit tests with the toy model used in \cref{sec:metric} one hundred times, either sampling data realizations from the likelihood at fixed input parameters ($\theta=\theta_0$, frequentist approach), or by first sampling the parameters from their priors ($\theta\sim{\rm P}(\theta)$, Bayesian approach). In both cases, we sample the parameter posterior distributions, generate the PPD and compute $\Delta_{\rm PPD}$, as illustrated in \cref{fig:toy-ppd2}. \Cref{fig:toy-ppd3} shows the histogram of $\Delta_{\rm PPD}$ values, and we find, in both cases, that it does not follow a uniform distribution (although these distributions are themselves indistinguishable using a two-sample Kolmogorov-Smirnov test).

\end{appendix}

\end{document}